\documentclass[reprint, superscriptaddress, amsmath,amssymb, aps, prb, floatfix, longbibliography]{revtex4-2}

\usepackage{amsfonts}
\usepackage{xcolor}
\usepackage{braket}
\usepackage{dsfont}
\usepackage{graphicx}
\usepackage[caption=false]{subfig}
\usepackage{listings}
\usepackage{comment}
\usepackage{hyperref}

\begin{document}
	
\preprint{APS/123-QED}
	
	\title{Majorana sweet spots in 3-site Kitaev chains}
    
	\author{Rodrigo A. Dourado}
	\affiliation{ \textit{Instituto de F\'isica de S\~ao Carlos, Universidade de S\~ao Paulo, 13560-970 S\~ao Carlos, S\~ao Paulo, Brazil}	}
    \author{Martin Leijnse}
    \affiliation{Division of Solid State Physics and NanoLund, Lund University, S-22100 Lund, Sweden}
	\author{Rub\'{e}n Seoane Souto}
    \affiliation{Instituto de Ciencia de Materiales de Madrid (ICMM), Consejo Superior de Investigaciones Cient\'{i}ficas (CSIC), Sor Juana In\'{e}s de la Cruz 3, 28049 Madrid, Spain}

	\date{\today}
	
\begin{abstract}
Minimal Kitaev chains, composed of two quantum dots (QDs) connected via a superconductor, have emerged as an attractive platform to realize Majorana bound states (MBSs). These excitations exist when the ground state is degenerate. The additional requirement of isolating the MBS wavefunctions further restricts the parameter space to discrete sweet spots. While scaling up to Kitaev chains with more than two sites has the potential to improve the stability of the MBSs, longer chains offer more features to optimize, including the MBS localization length and the excitation gap. In this work, we theoretically investigate 3-site Kitaev chains and show that there are three different types of sweet spots, obtained by maximizing distinct MBS properties: genuine 3-site sweet spots with well-localized MBSs at the ends, effective 2-site sweet spots, where the middle site acts as a barrier, and sweet spots with delocalized MBSs that overlap in the middle of the chain. These three cases feature different degrees of robustness against perturbations, with the genuine 3-site being the most stable. We analyze the energy spectrum, transport, and microwave absorption associated with these three cases, showing how to distinguish them.
\end{abstract}
	
	\maketitle

\section{Introduction}

Majorana bound states (MBSs) are non-local states predicted to appear at the end of one-dimensional (1D) topological superconductors, that feature a dominant equal-spin p-wave electron pairing \cite{leijnse2012introduction, alicea2012new, aguado2017, beenakker2020search, prada2020andreev, flensberg2021engineered, das2023search, kouwenhoven2024perspective}. The Kitaev model, {\it i.e.} a 1D tight-binding chain of spinless fermions that features p-wave pairing between neighboring sites, is the simplest theoretical model to study this topological phase~\cite{Kitaev2001}. The model predicts the emergence of MBSs exponentially localized at the ends of the chain for a range of parameters. In fact, there is a set of parameters for which the MBSs are fully localized on the outermost sites, which means that spatially separated MBSs can appear already in a minimal (2-site) Kitaev chain~\cite{Leijnse2012}.

Quantum dots (QDs) coupled via narrow superconductors is a promising system to emulate Kitaev chains~\cite{Sau2012, Fulga2013}. The superconductors mediate two types of indirect coupling between the QDs: elastic cotunneling (ECT) and crossed Andreev reflection (CAR). Under sufficiently strong spin-splitting in the QDs, two MBSs localize at the ends when CAR and ECT amplitudes are equal. These amplitudes are tunable in a setup where the intermediate superconductors between the QDs host subgap states, so-called Andreev bound states~\cite{liu2022tunable, tsintzis2022creating, liu2024enhancing}. In their minimal form, featuring only two normal QDs, these systems implement 2-site Kitaev chains~\cite{Leijnse2012, tsintzis2022creating}. Signatures consistent with MBSs have been measured in minimal Kitaev chains fabricated in nanowires~\cite{Dvir2023, bordin2023tunable, bordin2024crossed, zatelli2024robust} and 2D electron gases~\cite{ten2024two}. This progress motivated proposals for determining the localization of MBSs~\cite{Seoane2023,Alvarado_PRB2024}, demonstrating their non-Abelian signatures through fusion~\cite{liu2023fusion, pandey2024nontrivial} and braiding~\cite{tsintzis2024majorana, boross2024braiding}, and using them for qubits \cite{tsintzis2024majorana, pino2024minimal, pan2024rabi}. Other proposals involve directly coupling superconducting QDs or Josephson junctions with different phases~\cite{svensson2024quantum, luna2024flux, escribano2025phasecontrolledminimalkitaevchain}, normal-superconductor QD junctions~\cite{Miles2023}, and coupling normal QDs via floating superconducting islands~\cite{souto2024majorana}, see Ref.~\cite{Souto_arXiv24} for a recent overview.

However, in minimal Kitaev chains, MBSs exhibit limited protection, as parameter variations can cause their overlap or lift the ground-state degeneracy. Therefore, they are sometimes referred to as ``Poor man's Majoranas bound states''~\cite{Leijnse2012}. In this work, however, we will consistently use the term MBSs also for minimal 2-site Kitaev chains. Longer chains provide greater protection for MBSs~\cite{dourado2025twositekitaevsweetspots}, with a fully protected topological phase in the limit of an infinite chain. This notion motivated experiments on effective 3-site Kitaev chains, composed of three normal QDs coupled via two superconductors \cite{bordin2024signatures, Haaf_arXiv2024}, that demonstrated a higher degree of protection of the ground state degeneracy. However, depending on the physical model parameters and the properties of interest, 3-site chains do not always present an improvement over their 2-site counterparts \cite{svensson2024quantum}. 

\begin{figure}
 \includegraphics[width=1\linewidth]{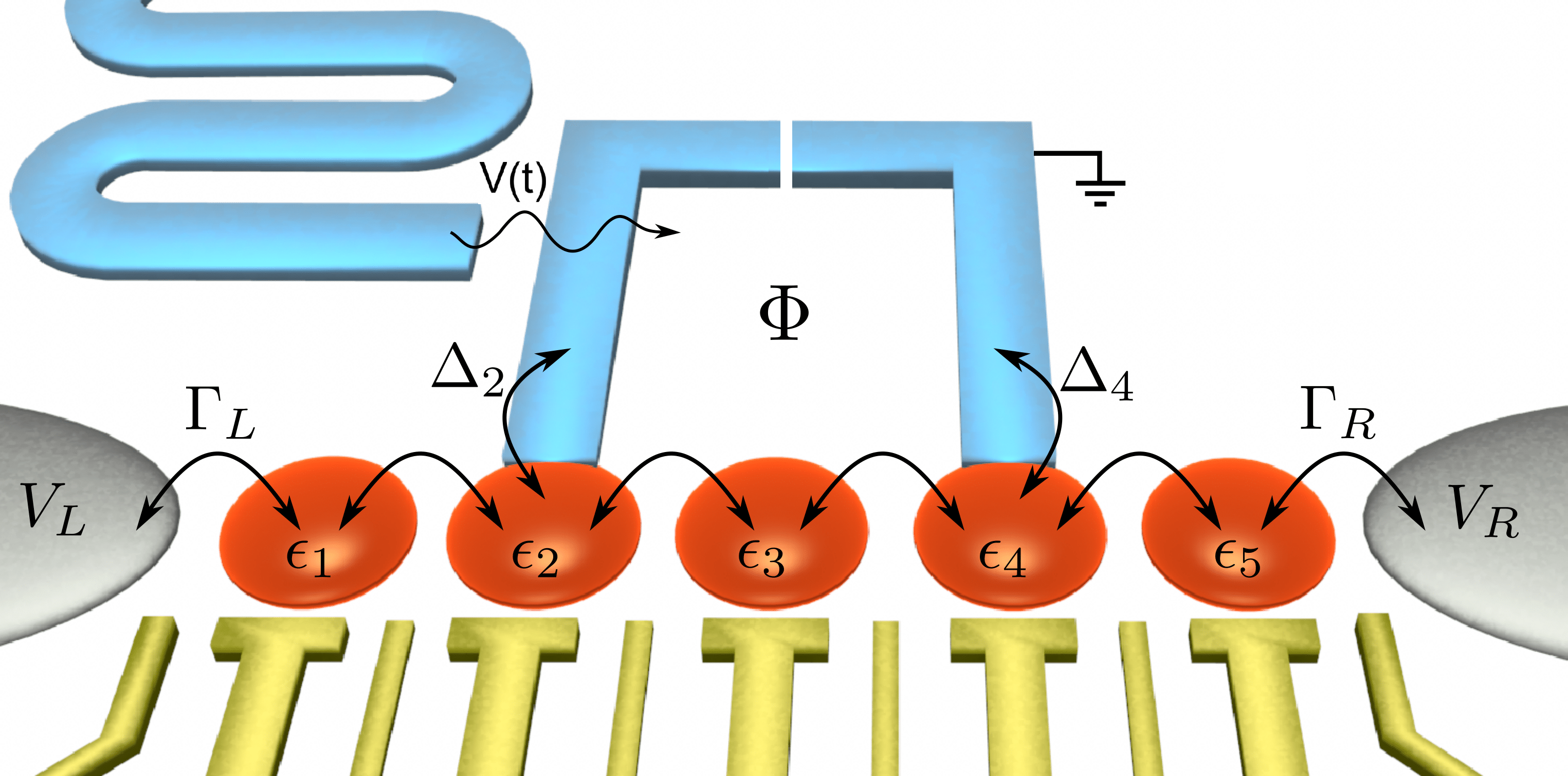}
  \caption{Sketch of the 3-site artificial Kitaev chain, where 5 QDs couple via tunneling. The even QDs couple to superconductors which gives rise to pairing potentials $\Delta_i$ ($i=2,4$). The system attaches to metallic leads to perform spectroscopy. The superconductors form a SQUID loop that allows for phase control and is capacitively coupled to a microwave resonator (top left).}
\label{Fig1}
\end{figure}

In this work, we theoretically analyze a QD-based 3-site Kitaev chain, as illustrated in Fig.~\ref{Fig1}. In the original Kitaev model, fully isolated zero-energy MBSs at the edges and a maximum excitation gap can be achieved concurrently at sweet spots in parameter space~\cite{Leijnse2012}. However, at finite magnetic fields, these features cannot be optimized simultaneously. Thus, the optimization of disctinct Majorana properties, such as isolated MBSs at the edges, quantified using the Majorana polarization (MP) \cite{sticlet2012spin, sedlmayr2015visualizing, sedlmayr2016majorana, aksenov2020strong}, and the excitation gap, for instance, leads to different types of sweet spots. From those, we highlight three representative cases: effective 2-site sweet spots, where the central QD is detuned, acting as a barrier that mediates CAR and ECT between the outermost QDs; genuine 3-site sweet spots with well-localized MBSs; 3-site sweet spots where the MBSs localize poorly, significantly overlapping inside the system, we refer to those as delocalized sweet spots for short. The genuine 3-site sweet spot is the ideal case, while the other sweet spots are reached by shifting the energies of the middle and the outermost QDs, respectively, while keeping ground state degeneracy.

The ground state splitting resulting from detuning some of the QDs allows us to distinguish between these three types of sweet spots. For instance, shifting the energy of both outermost QDs results in a quadratic energy splitting of the ground state in the effective 2-site sweet spot~\cite{Leijnse2012}, unlike for the other two cases, where the ground state remains degenerate. Shifting the energy of the central QD allows us to distinguish between the 3-site sweet spots with well-localized and delocalized MBSs, as the ground state remains degenerate only for the genuine 3-site sweet spot.
The phase difference between the superconductors is another parameter that can help distinguishing these three cases. We analyze the features as a function of the phase in both transport and microwave spectroscopy, discussing the unique features of genuine three-site sweet spot. 

The remainder of the text is organized as follows. We start discussing the model and formalism in Sec.~\ref{theoreticalModel}.We consider two different models, starting with the Kitaev model in Sec.~(\ref{resultsKitaev}), then switching to a microscopic model (\ref{reusltsMicroscopic}), that includes effects from a finite Zeeman splitting and charging energy in the QDs. For each case, we study the conductance and energy spectrum. In Sec. (\ref{phaseSection}), we explore the phase degree of freedom, investigating the conductance and microwave absorption spectrum. We present our concluding remarks in Sec. (\ref{conclusions}).

\section{Theoretical description} \label{theoreticalModel}

We consider a realization of the 3-site Kitaev chain using an array of $5$ QDs, as shown in Fig.~\ref{Fig1}. The QDs are made of semiconducting materials with large spin-orbit coupling (SOC). QDs $2$ and $4$ are proximitized by a grounded s-wave superconductor. The QD-superconductor coupling generates Andreev bound states localized on QDs $2$ and $4$ that mediate two types of processes between the normal QDs \cite{liu2022tunable, tsintzis2022creating}. In the ECT, single electrons are transferred between normal QDs, while the CAR represents the formation or splitting of Cooper pairs. Finally, to break time-reversal symmetry, we consider an applied magnetic field perpendicular to the direction of the SOC. The combination of applied magnetic field, s-wave superconductivity, and SOC opens up the possibility for equal-spin CAR between the QDs~\cite{wang2022singlet, Bordoloi_Nature2022}. The Hamiltonian that describes the system can be written as
\begin{equation} \label{fullModel Hamiltonian}
    H=H_c+H_l+H_{c-l},
\end{equation}
where $H_c=H_{\rm QDs}+H_T$ describes the artificial Kitaev chain, composed of five tunnel-coupled QDs. The QDs are described by
\begin{equation} \label{QDsH}
\begin{split}
    H_{\rm QDs} =& \sum_{i, \sigma} (\epsilon_{i, \sigma} + s_\sigma V_{z,i}) n_{i, \sigma} +\sum_i U_i n_{i, \uparrow}n_{i, \downarrow}+\\
    &\Delta_2 c_{2, \uparrow}^\dagger c_{2, \downarrow}^\dagger + \Delta_4c_{4, \uparrow}^\dagger c_{4, \downarrow}^\dagger + H.c.,
\end{split}
\end{equation}
where $\epsilon_{i, \sigma} = \alpha_i V_i$ is the on-site energy, tunable via an external gate voltage $V_i$ ($\alpha_i$ is the corresponding gate lever arm), $s_{\uparrow, \downarrow}= \pm1$, $V_{z,i}$ represents the Zeeman energy due to the applied magnetic field along the $z$-axis, and $n_{i, \sigma} = c_{i, \sigma}^\dagger c_{i, \sigma}$ is the electron number operator, where $c_{i, \sigma}^\dagger (c_{i, \sigma})$ creates (destroys) an electron with spin $\sigma$ on site $i$. The electron-electron repulsion, which is typically strong in QDs, is represented by $U_i$. We assume that QDs 2 and 4 are strongly coupled to the superconductor, which leads to effective superconducting pairings $\Delta_{2,4}$. The phase difference $\phi$ between the superconductors can be tuned if they are connected forming a loop, as sketched in Fig.~\ref{Fig1}. Moreover, the strong coupling to the superconductor significantly renormalizes the charging energy and the Zeeman splitting on the proximitized QDs. For this reason, we consider $U_2=U_4=0$ and $V_{z,2}=V_{z,4}=0$, although the qualitative results do not change if we relax these conditions.

The tunneling between the QDs is given by
\begin{equation} \label{TunnelH}
    H_T= \sum_{i, \sigma}\left[ t_i c_{i+1, \sigma}^\dagger c_{i, \sigma} + t_{i}^{so} s_{\sigma}c_{i+1, \sigma}^\dagger c_{i, \bar{\sigma}}+ H.c. \right],
\end{equation}
where $t_i$ ($t_i^{so}$) is the spin-conserving (spin-flipping) hopping, with the spin-orbit field along the $y$-axis. Here, $\bar{\sigma}$ denotes the opposite spin to $\sigma$.

The chain Hamiltonian $H_c$ can map onto the Kitaev model \cite{Kitaev2001} when we integrate out the degrees of freedom of the superconducting QDs $2$ and $4$. In this context, the ECT and CAR processes represent couplings between the normal QDs. In the presence of a sufficiently large magnetic field, the effective low-energy Hamiltonian reads \cite{liu2022tunable}
\begin{equation} \label{HKitaev}
    H_K=\sum_{j=1}^3 \epsilon_j \Tilde{c}_j^\dagger \Tilde{c}_j+\sum_{j=1}^{2}\left(\tau_j \Tilde{c}_j^\dagger \Tilde{c}_{j+1}+\delta_j \Tilde{c}_j^\dagger \Tilde{c}^\dagger_{j+1}+{\rm H.c.}\right),
\end{equation}
where $\Tilde{c}_j$ is a combination of $c_{j, \uparrow}$ and $c_{j, \downarrow}$. Here, $\tau_j$ is the hopping, associated with ECT processes, and $\delta_j$ represents the effective p-wave superconducting pairing amplitude, given by the CAR amplitude between QDs without proximity effect.

Both the microscopic model [Eqs.~(\ref{QDsH}-\ref{TunnelH})] and the Kitaev model [Eq.~(\ref{HKitaev})] predict the existence of even-odd ground state degeneracies, where $\delta E_0 = E_0^{odd} - E_0^{even}=0$, with $E_0^{even(odd)} \ket{E (O)}$ being the lowest energy eigenstate with even (odd) fermion parity. To characterize the sweet spots, we use the MP calculated at the ends of the chain~\cite{sticlet2012spin, sedlmayr2015visualizing, tsintzis2022creating, Awoga_PRB24}
\begin{equation} \label{MP eq}
    M_j = \frac{\sum_{\sigma} w_{j, \sigma}^2 - z_{j, \sigma}^2}{\sum_\sigma w_{j, \sigma}^2 + z_{j, \sigma}},
\end{equation}
that takes a value $\pm 1$ whenever only one MBS has weight on site $j$ \footnote{The above equation is valid for real Hamiltonians. For $\phi \neq 0$ this is no longer the case, and phase rotations in the $c_{j, \sigma}$ operators are needed to maximize the MP, see Ref.~\cite{svensson2024quantum}}. Here, $w_{j, \sigma} = \bra{O}( c_{j, \sigma} + c_{j, \sigma}^\dagger ) \ket{E}$ and $z_{j, \sigma} = \bra{O}( c_{j, \sigma} - c_{j, \sigma}^\dagger ) \ket{E}$, which are the MBS wavefunctions coefficients $\gamma_{j,\sigma}^1$ and $\gamma_{j,\sigma}^2$, respectively. Note that the MP does not contain direct information about the MBS localization length or the MBS overlap in the middle of the chain.

We couple the system to metallic leads, whose Hamiltonian is
\begin{equation}
    H_l = \sum_{k, \sigma,\alpha} \left( \Omega_k - \mu_{\alpha} \right) d_{k, \sigma, \alpha}^\dagger d_{k, \sigma, \alpha},
\end{equation}
where $\alpha = L, R$ labels the lead, $k$ is the electron momentum, which gives the kinetic energy $\Omega_k = k^2/2m$ ($e = k_B = \hbar = 1$). The tunneling Hamiltonian that couples chain and leads is
\begin{equation}
    H_{c-l} = \sum_{k, \sigma} \left[t_{L} d_{k, \sigma, L}^\dagger c_{1, \sigma} + t_R d_{k, \sigma, R}^\dagger c_{5, \sigma} +H.c. \right]. 
\end{equation}

For the microscopic model, Eqs.~(\ref{QDsH}-\ref{TunnelH}), we calculate transport properties through the system using a rate equation approach. In this framework, we solve equations for the reduced density matrix in the regime $\Gamma_\alpha = 2\pi \rho t_\alpha^2 \ll T$, where $\rho$ is the density of states of the leads, using the QMEQ python package \cite{kirvsanskas2017qmeq,tsintzis2022creating}. The Kitaev model is exactly solvable in the single-particle picture, allowing us to calculate transport using scattering matrix theory. We define the conductance as
\begin{equation}
    G_{\alpha\beta}=\frac{d I_\alpha}{d V_\beta}\,,
\end{equation}
with $\alpha,\beta=L,R$, where $\alpha=\beta$ and $\alpha\neq\beta$ correspond to the local and nonlocal conductance, respectively.

In addition to the conductance, we analyze the microwave absorption spectrum of the Kitaev chains. We consider the superconductors composing a superconducting quantum interference device (SQUID) loop, where the left superconductor is capacitively coupled to a resonator, as depicted in Fig.~\ref{Fig1}, see Ref.~\cite{zhuo2025readfermionparitypotential} for an implementation in a minimal Kitaev chain. The resonator induces a small time-dependent voltage on the superconductor, $V(t) \propto \cos(\omega t)$, where $\omega$ characterizes the resonator's frequency. Due to the Josephson relation, $\frac{d \delta \phi (t)}{dt} = 2 V(t)$, this voltage causes a time-dependent perturbation of the phase difference between the superconductors~\cite{vayrynen2015microwave}. Transitions to excited states occur when $\omega$ matches the energy difference between two states with the same fermion parity~\cite{vayrynen2015microwave}. In linear response, the absorption spectrum can be expressed as $S(\omega) = \sum_m |\bra{m}\hat{n}\ket{0}|^2 \delta(\omega - \omega_{0m})$, where $\hat{n} = \partial/\partial \hat{\phi}$ is the Cooper pair number operator, with $\hat \phi$ being the phase operator, $\ket{0}$ is the ground state, and $m$ labels the excited states~\cite{pino2024minimal}. In addition, $\omega_{0m} = \omega_m - \omega_0$ represent the resonance peaks, which are weighted by the transition elements to states within the same parity sector.

\section{Results}

In this section, we discuss the results for the two presented models, focusing on the different types of sweet spots that appear. 

\subsection{Kitaev chain} \label{resultsKitaev}

\begin{figure*}
\centering
\includegraphics[width=1\linewidth]{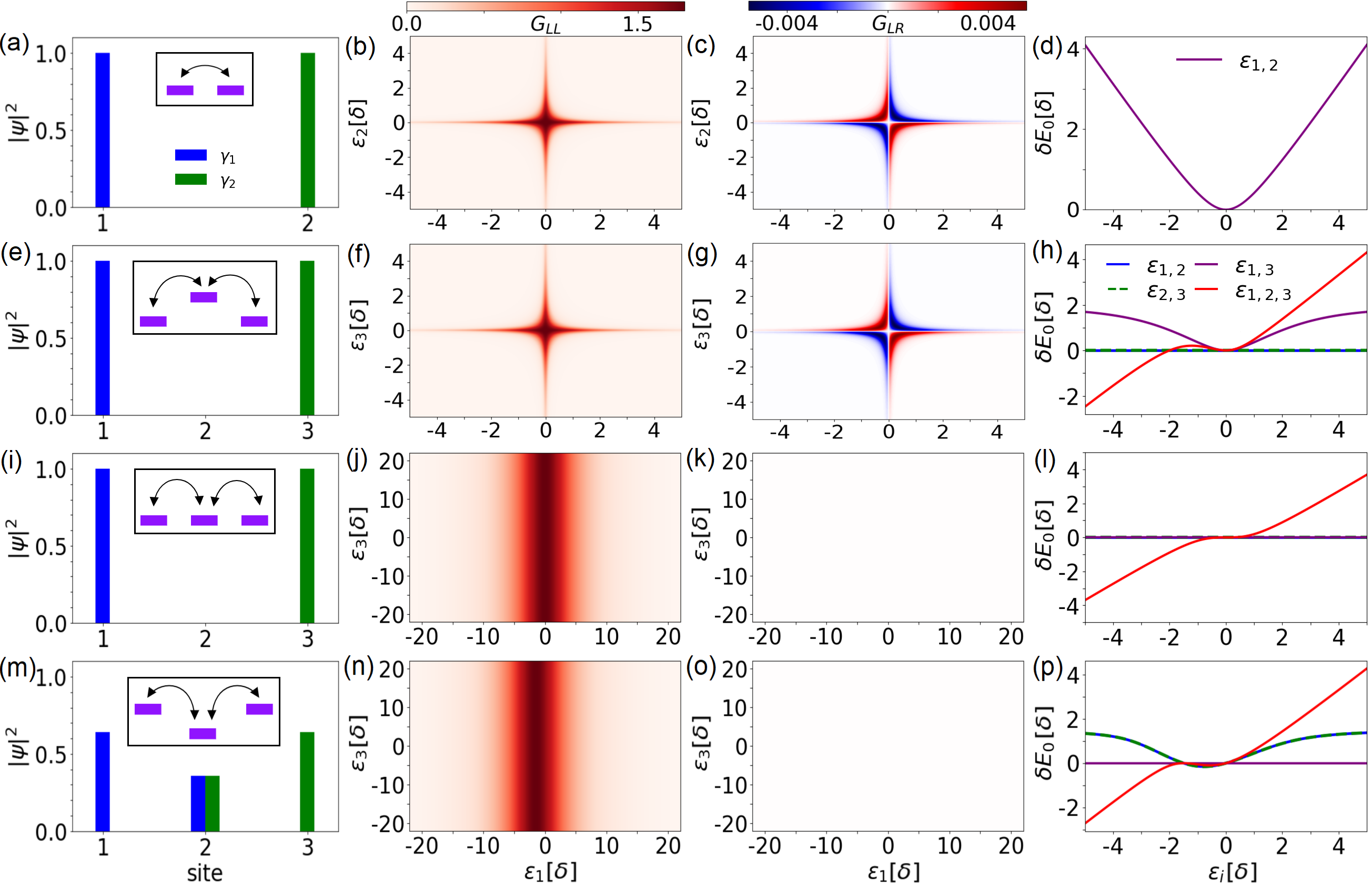}
  \caption{Conductances and energies for the Kitaev model. (a-d): true 2-site Kitaev chain. (e-h): effective 2-site sweet spot in a 3-site chain, where the middle site is detuned by $2\delta$ and acts as a barrier. (i-l): 3-site sweet spot, $\epsilon_{1, 2, 3} = 0$. (m)-(p): 3-site sweet spot with the outer sites detuned by $1.5\delta$. In all cases, $\tau = \delta$. In the first to fourth columns we show the wavefunctions, local conductance, nonlocal conductance, and $\delta E_0$, respectively. For the conductance calculations, we set $T = \delta/100$ and $\Gamma_L = \Gamma_R = 0.1 \delta$.}
\label{Fig2}
\end{figure*}

We begin our analysis by studying the Kitaev model~\eqref{HKitaev}, which hosts two perfectly localized MBSs at the ends ($|M_1|=|M_N|=1$) at discrete sweet spots, $\tau_i=\delta_i$ and $\epsilon_i = 0$, regardless of the chain length ($N\geq2$).

In minimal Kitaev chains ($N=2$), Figs. \ref{Fig2}(a-d), the ground state degeneracy does not split when one of the site energies is detuned away from the sweet spot \cite{Leijnse2012}. This results in a peak in the local conductance, Fig.~\ref{Fig2}(b), and a zero in the nonlocal conductance at zero applied bias, Fig.~\ref{Fig2}(c), that remains when either $\epsilon_1=0$ or $\epsilon_2=0$. The ground state splits when two site energies are detuned, showing a quadratic dependence with the level energies, Fig.~\ref{Fig2}(d). This results in a decrease of the local conductance and an increase in the nonlocal conductance with a sign that depends on whether CAR or ECT dominates~\cite{tsintzis2022creating}.

The 3-site Kitaev model features localized MBSs for $\tau_i = \delta_i$ and $\epsilon_i = 0$. Notably, there is robustness to parameter variations, resulting in extended regions of parameter space where the sweet spot conditions, i.e. degeneracy of the ground state and high MP, are satisfied. Among these, we identify three illustrative cases that provide insight into the distinct types of sweet spots achievable in the microscopic model through optimization of the MBS properties, as we shall discuss in Sec.~\ref{reusltsMicroscopic}: an effective 2-site sweet spot, characterized by the middle site acting as a barrier; a genuine 3-site sweet spot with localized MBSs at sites $1$ and $3$; a 3-site sweet spot where $\epsilon_1$ and $\epsilon_3$ are detuned, causing the MBSs to overlap at the middle site. We refer to the latter as the delocalized sweet spot.

We illustrate the results for the effective 2-site sweet spot in Figs.~\ref{Fig2}(e-h). In this case, the central site is detuned, $|\epsilon_2| \gg |\tau|,|\delta|$, forming a barrier between the outermost sites, that causes the 3-site Kitaev chain to behave as an effective 2-site chain. The barrier leads to effective CAR and ECT amplitudes between the outermost sites that are smaller than the corresponding bare amplitudes. Consequently, the excitation gap is reduced below $2\tau = 2\delta$, the value for the sweet spot for the true two-site Kitaev chain. In this regime, the local and nonlocal conductance features resemble the ones of the 2-site sweet spot previously discussed \cite{tsintzis2022creating, Dvir2023, ten2024two}, compare the first and second rows in Fig.~\ref{Fig2}. In both cases, the ground state splits quadratically when detuning the outermost sites. Effective 2-site sweet spots in 3-site Kitaev chains exhibit robustness against detunings of site energies, provided these detunings do not simultaneously involve sites $1$ and $3$, see Fig.~\ref{Fig2}(h). Additionally, setting $\tau_{1(2)} = \delta_{1(2)}$ and $\epsilon_{1(3)}=0$ results in a disconnected MBS on the left (right) side, ensuring ground state degeneracy regardless of the value of $\tau$ or $\delta$ on the other side.

The {\it genuine} 3-site sweet spot appears for $\tau=\delta$ and $\epsilon_j=0$ ($j=1,2,3$), see Figs.~\ref{Fig2}(i-l). In this case, detuning one of the outermost sites converts the 3-site into an effective 2-site sweet spot. This results in the local conductance being independent of the energy of the site at the opposite end, see Fig.~\ref{Fig2}(j). Additionally, in the 3-site sweet spots, the wavefunctions of the MBSs shift away from the edges more gradually as $\epsilon_{1,3}$ increase, compared to the effective 2-site case. In the latter, changes in $\epsilon_{1,3}$ have a more pronounced effect on the MBSs due to the reduced effective pairing amplitude between the outermost sites in the effective 2-site description, which is smaller than the original gap parameter $\delta$ of the 3-site chain. This slower shift results in the local conductance remaining stable over a broader range of $\epsilon_1$ in Figs.~\ref{Fig2}(j), compared to Fig.~\ref{Fig2}(f). The nonlocal conductance remains fully suppressed at the 3-site sweet spot, Fig.~\ref{Fig2}(k), when sites $1$ and $3$ are detuned, in contrast to the effective and true 2-site sweet spots, Figs.~\ref{Fig2}(c, g). The ground state remains degenerate when detuning one or two sites simultaneously. Breaking the degeneracy requires detuning all 3 sites away from the sweet spot condition, causing the nonlocal conductance to acquire a finite value. Even then, the energy remains closer to zero, in comparison to the effective 2-site case, as the ground state splits cubically for a range 
of $|\delta \epsilon_{1, 2, 3}| \lesssim \delta$, signaling the increase in protection, see Fig.~\ref{Fig2}(l). 

The results discussed above show that the local and nonlocal conductances can be used to distinguish between the 2- and 3-site sweet spots. In reality, the extra features found in Figs.~\ref{Fig2}(f-g) with respect to Figs.~\ref{Fig2}(j-k) progressively appear as the middle site is detuned from the 3-site sweet spot. Additionally, removing a $|\epsilon_2| > 0$ barrier significantly increases the protection of this system, as MBSs can leak into the middle site without lifting the ground state degeneracy or reducing MP. For instance, for a 3-site sweet spot, when we vary $\epsilon_1$ or $\epsilon_3$, the MBS wavefunction is increasingly pushed into site $2$, while keeping the ground state degeneracy. 

The robustness of the ground state degeneracy to variations of the energy of the outer sites enables a new set of 3-site sweet spots where the MBS wavefunctions are pushed toward the middle site, decreasing the weight in the outermost part of the chain, see Fig.~\ref{Fig2}(m). These sweet spots show qualitatively the same transport signatures as the 3-site ones with localized MBSs, compare Figs.~\ref{Fig2}(n-o) and Figs.~\ref{Fig2}(j-k). We note that a lead attached to the central site can be used to measure the wavefunction localization, as done in Ref.~\cite{Haaf_arXiv2024}. At the spectroscopy level, the main difference is the splitting of the ground state degeneracy when the middle site is detuned, as shown in Fig.~\ref{Fig2}(p). 

The sweet spot shown in Fig.~\ref{Fig2}(m) illustrates the connection between robustness and localization of the MBSs. For example, since the MBSs do not overlap in sites $1$ and $3$, the ground state degeneracy remains robust against changes in $\epsilon_1$ and $\epsilon_3$. Conversely, the overlap of the MBSs at site $2$ leads to a linear splitting of the degeneracy when changing $\epsilon_2$, with the slope determined by the weight of the MBSs in site $2$ (not shown). In general, the localization of the MBSs can be assessed by detuning one or multiple QDs, providing insight into the robustness of the ground state degeneracy.

\subsection{Microscopic model} \label{reusltsMicroscopic}

After investigating the Kitaev model and introducing the different types of sweet spots, we move to the microscopic model given by Eqs.~(\ref{QDsH}-\ref{TunnelH}), which includes explicitly the Andreev states that mediate the coupling between the normal QDs, as well as the finite Zeeman splitting and charging energy of the QDs. \cite{tsintzis2022creating}. The microscopic model converges to the Kitaev chain in the limit $t \ll \Delta, V_z$. In this section, we consider the system's parameters to have left-right symmetry, so $\epsilon_1=\epsilon_5$, $\epsilon_2=\epsilon_4$, $|\Delta_2|=|\Delta_4| = \Delta$, $t_i=t_{5-i}$, and $t^{so}_i=t^{so}_{5-i}$ for $i=1,2$. This is a convenient choice to reduce the number of parameters, but it is not a requirement for finding sweet spots \cite{bordin2024signatures}. In the following, we choose $U = 5\Delta$, $V_z = 2.5\Delta$, $t_i = 0.5\Delta$, $t_i^{so} = 0.2t$, $T = \Delta/100$ and $\Gamma_L = \Gamma_R = T/10$.

\begin{figure}
\centering
\includegraphics[width=1\linewidth]{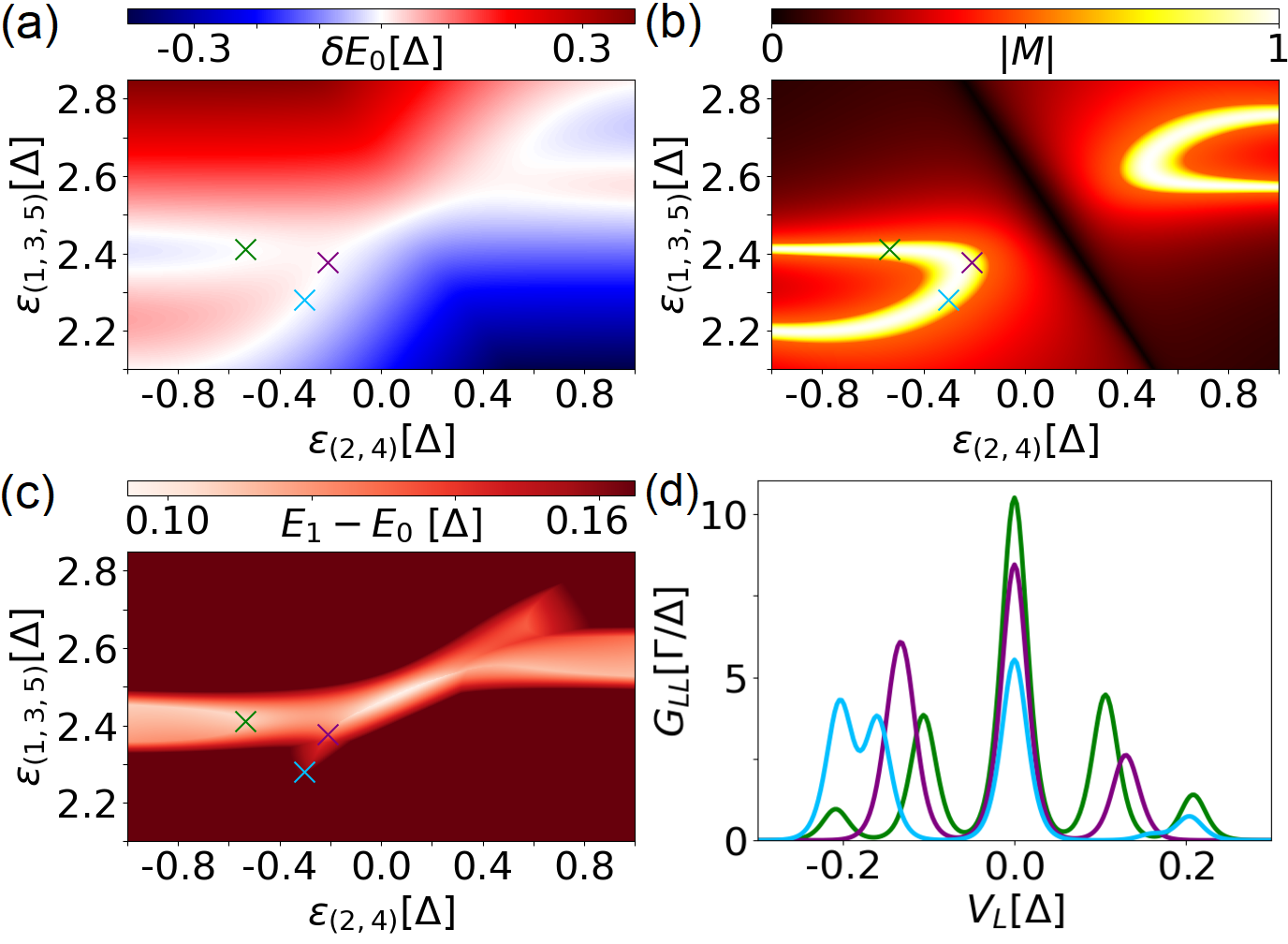}
\caption{(a) Energy difference between the even and odd ground states, $\delta E_0= E_0^{odd} - E_0^{even}$, (b) MP, and (c) excitation gap, as functions QD levels. (d) Local conductance at the sweet spots, indicated by the crosses in panels~(a-c) as a function of the voltage bias.}
\label{Fig3}
\end{figure}

\begin{figure*}[htb!]
\centering
\includegraphics[width=1\linewidth]{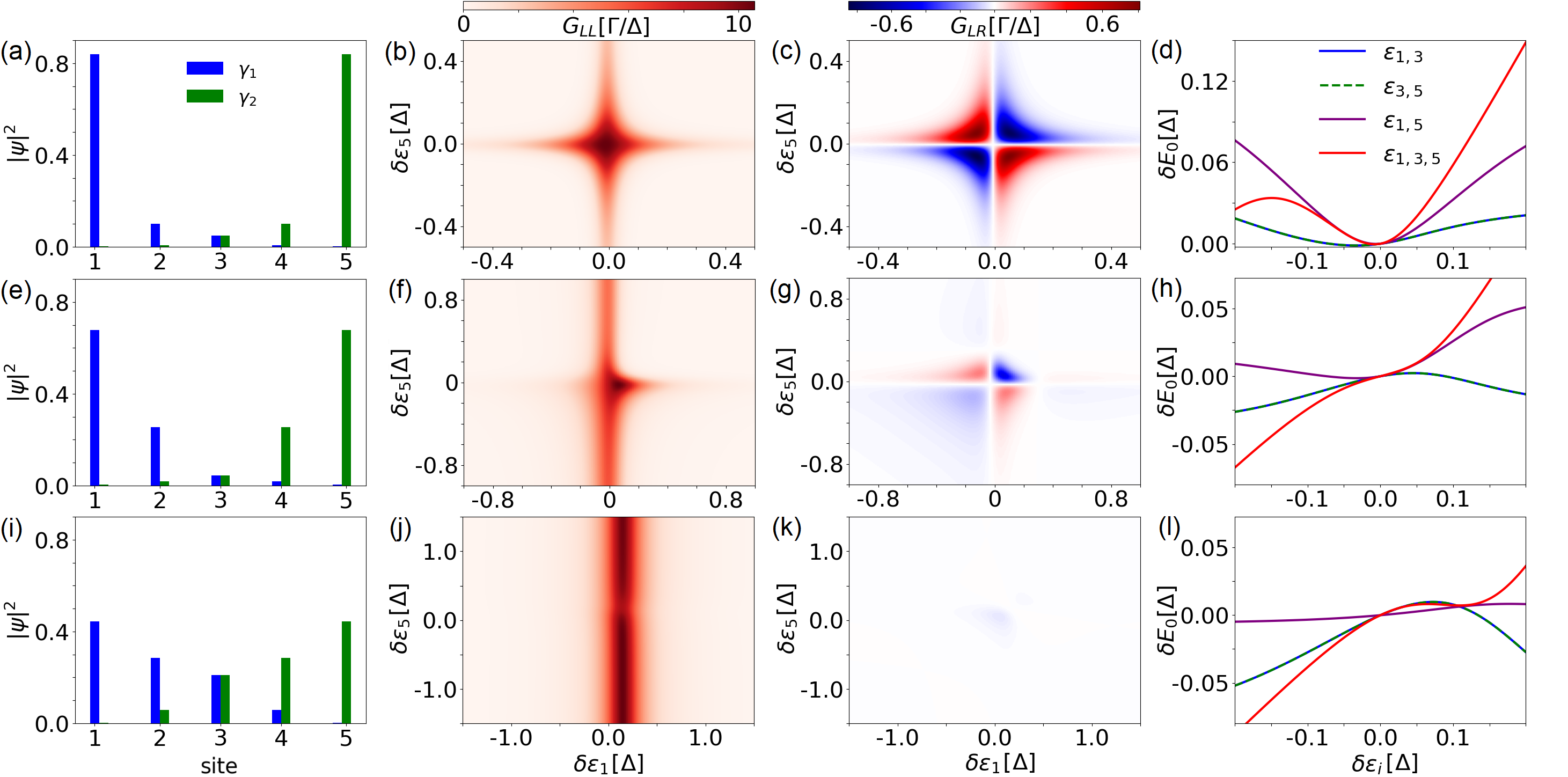}
\caption{Illustrative sweet spots for the microscopic model. The first to third rows show the results for the green, purple, and cyan sweet spots shown in Fig.~\ref{Fig3}, which are ordered with increasing values of the excitation gap. (a, e, i) present the wavefunctions of the MBSs. (b, f, j) and (c, g, k) show the local and nonlocal conductances, respectively. (d, h, l) show the dependence of $\delta E_0$ with respect to variations of multiple QD levels $\epsilon_i$.}
\label{Fig4}
\end{figure*}

We start by analyzing the parameter space to find zero-energy solutions with high MP at the ends of the QD array. We show results for the ground state splitting in Fig.~\ref{Fig3}(a), where the red/blue color indicates an even/odd fermion parity ground state. In between (white color), the ground state is degenerate, the first sweet spot condition. The MP, $|M_1| = |M_5| = |M|$, Fig.~\ref{Fig3}(b), peaks at values close to unity for some parameters. Unlike in the Kitaev model, the microscopic model with finite $V_z$ does not host fully isolated MBSs at the ends. 

The excitation gap, defined as $E_1 - E_0$, where both energies are taken within the lowest-energy parity sector, significantly varies along the $\delta E_0 = 0$ lines in Fig.~\ref{Fig3}(a). One reason for this variation is a misalignment of the chemical potential of the middle QD, causing it to act as a barrier, thereby reducing the excitation gap. The formation of inner barriers within the array of QDs can occur even though we keep $\epsilon_1=\epsilon_3=\epsilon_5$ because of the different renormalization effects experienced by the central normal QD, coupled to two superconductors, unlike the outermost QDs that couple to only one.

Here, we follow the white lines in Fig.~\ref{Fig3}(a), optimizing different characteristics such as the MP and excitation gap, to identify various types of sweet spots. We then select three representative sweet spots: the one with the highest MP, one with well-localized MBSs while maintaining MP and excitation gap values above a minimal threshold, and the one with the largest gap. These sweet spots, indicated by the green, cyan, and purple crosses in Figs. \ref{Fig3}(a-c), correspond to the types identified in our previous analysis (Fig.~\ref{Fig2}), and the excitation gap for each case is shown in Fig.~\ref{Fig3}(d) through the local conductance.

First, we analyze the sweet spot indicated by the green crosses in Figs.~\ref{Fig3}(a-c). The MBS wavefunctions are well-localized on QDs $1$ and $5$, with no overlap between the MBSs at the ends, Fig.~\ref{Fig4}(a). The local, Fig.~\ref{Fig4}(b), and nonlocal, Fig.~\ref{Fig4}(c), conductances are consistent with the signatures of an effective 2-site Kitaev chain, see Figs.~\ref{Fig2}(a-g) for comparison. Here, the ground state degeneracy is lifted when the outer QDs are detuned, Fig.~\ref{Fig4}(d), which explains both the suppression of $G_{LL}$ and the increase in $G_{LR}$ away from the sweet spot condition. These results indicate the formation of a barrier inside the system, allowing us to identify this case as an effective 2-site sweet spot.

We next focus on a sweet spot with a larger excitation gap, represented in purple in Figs.~\ref{Fig3}(a-d), which signals that the central QD does not act as a barrier to the same extent as in the effective 2-site sweet spot discussed above. The MBSs in this case show only a very small overlap in the central region, see Fig.~\ref{Fig4}(e), which is also the case for the effective 2-site sweet spot, Fig.~\ref{Fig4}(a). In contrast to the previous case, see Figs.~\ref{Fig4}(a-d), some of the 2-site characteristic conductance features are fainter. For instance, the local conductance, Fig.~\ref{Fig4}(f), mostly depends on the level of the QD that the lead is attached to, $\delta \epsilon_1$, while being almost independent of the level of the QD at the opposite side, $\delta \epsilon_5$. Additionally, the nonlocal conductance, Fig.~\ref{Fig4}(g), is suppressed compared to Fig.~\ref{Fig4}(c). Furthermore, in comparison to Fig.~\ref{Fig4}(d) the ground state shows a less pronounced splitting when detuning the outer QDs, and splits cubically when the three normal QDs are detuned, see Fig.~\ref{Fig4}(h), signaling enhanced protection. This situation resembles the genuine 3-site sweet spot from Kitaev model, see Figs.~\ref{Fig2}(i-l), and will henceforth be referred to as such. We note, however, that the energy splitting due to variations in $\epsilon_{1, 5}$ and consequently the finite nonlocal conductance indicates that this case does not perfectly map onto the genuine 3-site sweet spot in the Kitaev model, a problem that could be partially solved with further optimizations considering $\epsilon_3$ as a free parameter. 

Finally, we look for the sweet spot with the largest excitation gap, depicted by the cyan color in Figs.~\ref{Fig3}(a-d). The transport and ground state properties are consistent with a 3-site sweet spot as mentioned above, see Figs.~\ref{Fig2}(m-p). The delocalization of the wavefunctions leads to MBS overlap in the central QDs rendering them fragile against gate voltage fluctuations in the middle QD. Thus, this case can be distinguished from the previous two scenarios due to pronounced energy splittings in gate detunings involving $\epsilon_3$, as shown in Fig.~\ref{Fig4}(l). In this case, the optimization required to avoid the central QD acting as a barrier causes the detuning of the renormalized levels on the outermost QDs, similar to the delocalized 3-site sweet spot in Fig.~\ref{Fig2}(m). This pushes the MBSs wavefunctions toward the middle of the system, resulting in a significant MBS overlap, Fig.~\ref{Fig4}(i).

It is worth mentioning that, in reality, it is hard to make a sharp distinction between the three illustrated cases, as the system features sweet spots that can show intermediate behaviors between them. To identify optimal sweet spots for a given value of $V_z$, each parameter must be individually tuned to optimize desirable features, such as ground state degeneracy, maximum MP, excitation gap, and MBS localization. This optimization procedure is well-suited to machine learning techniques~\cite{Koch_PRAp2023, van2024cross, Benestad_PRB2024}, while we choose to focus on the described representative scenarios for illustrative purposes.

\subsection{Phase control} \label{phaseSection}

\begin{figure}
\centering
\includegraphics[width=1\linewidth]{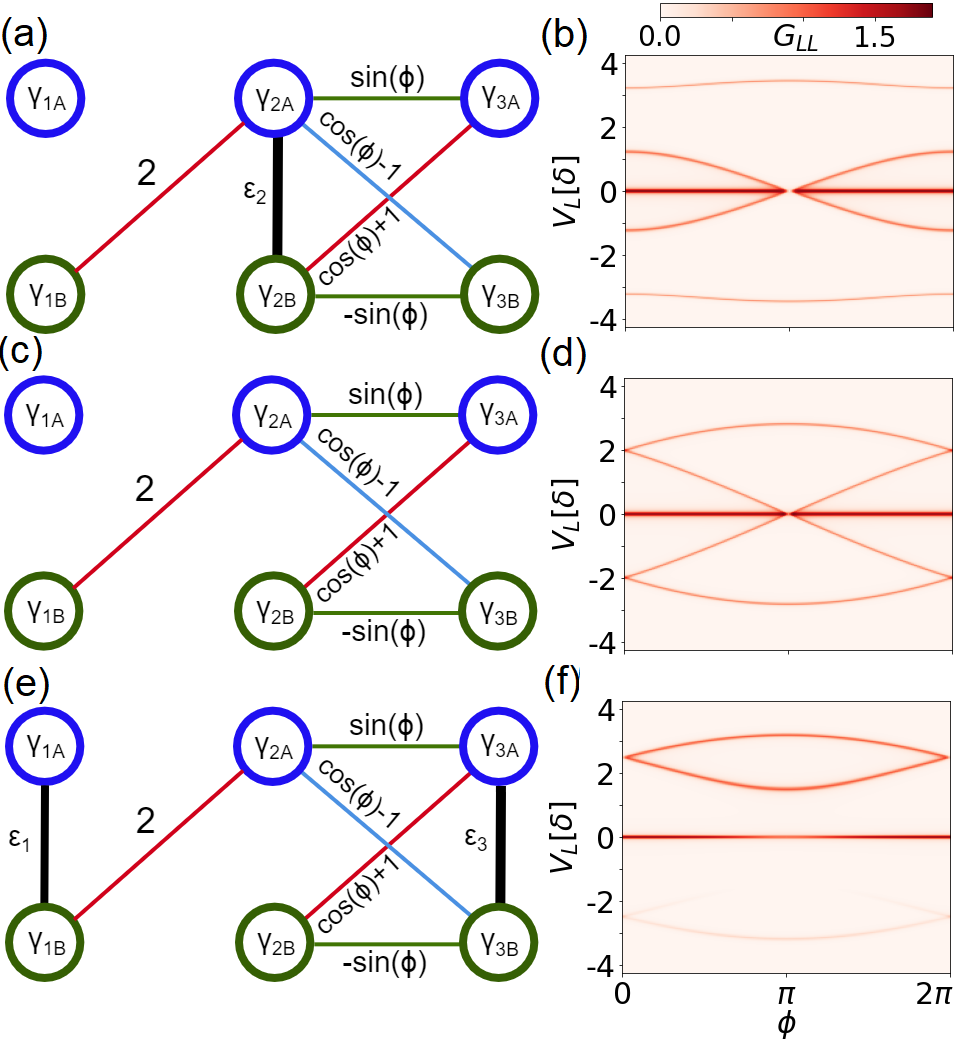}
\caption{Phase dependence of the 3-site Kitaev chain. (a), (c), and (e) diagrams showing the couplings between the MBSs as a function of $\phi$ for the three typoe of sweet spots. (b), (d), and (f) local conductance as a function of $\phi$ and bias voltage $V_L$, revealing the energy spectrum. In the first to last rows, we show the effective 2-site sweet spot, the genuine 3-site sweet spot, and the delocalized sweet spot, respectively.}
\label{Fig5}
\end{figure}

\begin{figure}
\centering \includegraphics[width=\linewidth]{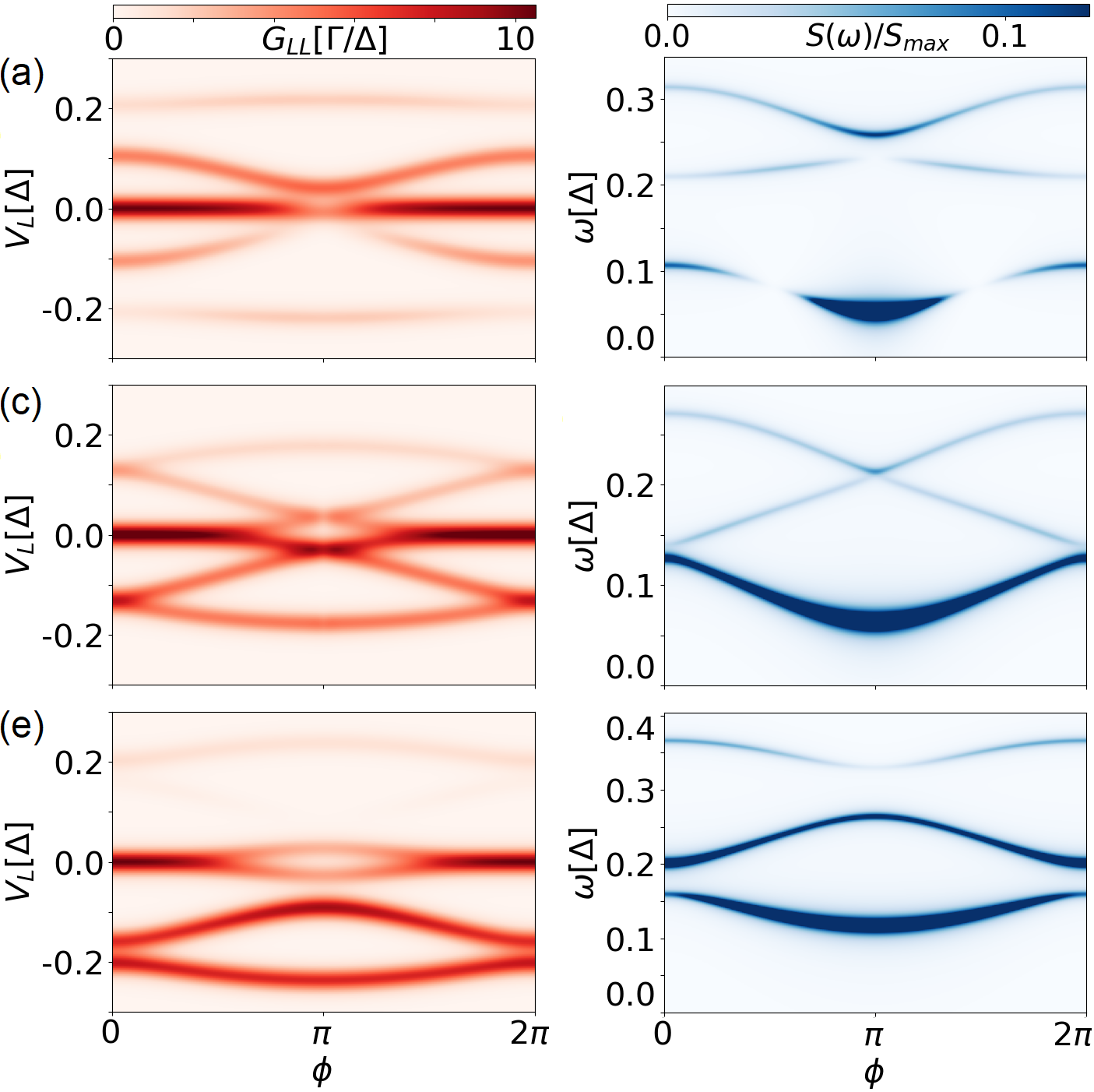}
\caption{(a), (c), and (e) Local conductance as a function of the voltage applied to the left normal lead and the phase difference between the superconductors. (b), (d), and (f) Microwave absorption spectrum for the lowest-energy (odd) parity sector as a function of $\phi$. The first to third rows show the results for the effective 2-site sweet spot, the genuine 3-site sweet spot, and the delocalized sweet spot highlighted in Fig.~\ref{Fig3}, respectively.}
\label{Fig6}
\end{figure}

Now we discuss the effect of the phase difference between the superconductors, investigated experimentally in Ref.~\cite{Haaf_arXiv2024}. We start with the Kitaev model~\eqref{HKitaev} to gain intuition. Performing the substitution of the MBS operators $c_j = \frac{1}{2} (\gamma_{jB} + i \gamma_{j A})$, the Kitaev Hamiltonian for 3 sites, Eq.~\eqref{HKitaev}, can be written as
\begin{equation}
\begin{split}
\label{Eq:Kitaev_transformed}
    &H_K = \frac{i}{2} \left\{ \sum_{j=1}^3 \epsilon_j \gamma_{jB}\gamma_{jA} - 
2\gamma_{1B}\gamma_{2A} +\gamma_{2A}\gamma_{3B}[\cos(\phi) - 1] \right. \\
& \left. \vphantom{\sum_i} + \gamma_{2B}\gamma_{3A}[\cos(\phi) + 1] - (\gamma_{2A}\gamma_{3A} - \gamma_{2B}\gamma_{3B}) \sin(\phi)  \right\},
\end{split}
\end{equation}
where we have taken $|\tau| = |\delta|$ at both sides of the system. We have chosen a gauge in which the superconducting phase is included in the p-wave pairing potential at the right side. This choice simplifies the interpretation of the MBS couplings and does not affect the results. Equation~\ref{Eq:Kitaev_transformed} illustrates that detuning the onsite energy, $\epsilon_i\neq0$, couples the local MBS operators on site $i$, black lines in the diagrams of Figs.~\ref{Fig5}(a, c, e). 

We start with the simplest case, the genuine 3-site sweet spot, where all $\epsilon_i = 0$, Fig.~\ref{Fig5}(c). For $\phi=0$, two MBSs localize at the ends of the chain, $\gamma_{1A}$ and $\gamma_{3B}$, leading to a degenerate ground state and giving rise to the zero-bias conductance peak in Fig~\ref{Fig5}(d). In addition, the other four MBS operators can be combined to form two fermions, $\gamma_{1B} + i\gamma_{2A}$ and $\gamma_{2B} + i \gamma_{3A}$, which are degenerate with energy $2t = 2\delta$, as shown in Fig.~\ref{Fig5}(d). For a general superconducting phase, $\gamma_{1A}$ remains disconnected, while the other 5 MBS operators have $\phi$-dependent couplings, see Eq.~\eqref{Eq:Kitaev_transformed} and the sketch in Fig.~\ref{Fig5}(c). This means that, out of these 5 MBS operators, 2 conventional fermions form while keeping 1 MBS uncoupled, which ensures degeneracy of the ground state for any $\phi$, a feature observed for all three sweet spots. However, the phase reduces the gap to the lowest excited state, which goes to zero for $\phi=\pi$, where the ground state is four-fold degenerate. The origin of this degeneracy can be understood by looking at the sketch in Fig.~\ref{Fig5}(c): for $\phi=\pi$ there are 3 decoupled MBSs, $\gamma_{1A}$, $\gamma_{2B}$, and $\gamma_{3A}$. The three remaining MBS operators form a finite-energy fermion and a delocalized MBS \cite{Haaf_arXiv2024}.

Forming a barrier between the outer sites, {\it i.e.} detuning $\epsilon_2$, couples the MBS operators at the central site, Figs.~\ref{Fig5}(a-b). This coupling splits the excited state at $\phi=0$, Fig. 5(b), providing a way of distinguishing between the effective 2-site and the genuine 3-site sweet spots. The excitation gap is also modified by $\phi$. For $\phi = \pi$, $\gamma_{1A}$ and $\gamma_{3A}$ are uncoupled similar to the genuine 3-site sweet spot. However, $\gamma_{2A}$ now couples to three MBS operators, $\gamma_{1B}, \gamma_{2B}$, and $\gamma_{3B}$. Interestingly, although there is a four-fold degeneracy, the zero-bias conductance vanishes near $\phi = \pi$. The reason for this is that $\gamma_{2A}$ and $\gamma_{2B}$ compose a fermionic state with large energy, such that $\gamma_{1B}$ and $\gamma_{3B}$ become effectively decoupled. Therefore, localized fermionic states emerge in sites $1$ and $3$, causing $G_{LL}$ to vanish due to interference effects from having both MBSs coupled to the same lead.

The delocalized 3-site sweet spot arises when the outer QDs are detuned, which we indicate in Fig.~\ref{Fig5}(e) by the black lines for $\epsilon_1$ and $\epsilon_3$. In this case, there are two sets of decoupled MBS operators: $\{\gamma_{1A}, \gamma_{1B}, \gamma_{2A}\}$ and $\{\gamma_{2B}, \gamma_{3A}, \gamma_{3B}\}$, Fig.~\ref{Fig5}(e), from which delocalized MBSs emerge. The odd number of MBS operators on each side ensures two decoupled MBSs. The symmetry between the two subsets of operators gives rise to a degeneracy between the second and third excited states at $\phi = 0$, as shown in Fig.~\ref{Fig5}(f) (similar to the genuine 3-site sweet spot). For $\phi > 0$ the degeneracy of the fermionic modes splits due to the local coupling between the MBSs, but the energy of the lowest mode sticks to zero. Finally, for $\phi = \pi$ there is no four-fold degeneracy as in the previous cases. Instead, we observe a decoupled MBS operator in the middle site, $\gamma_{2B}$, while the $5$ other operators are now connected due to $\epsilon_1$ and $\epsilon_3$. This leads to an avoided crossing between the two lowest modes and the ground state remains doubly degenerate. 

In summary, detunings from the genuine 3-site sweet spot, Fig.~\ref{Fig5}(d), lead to energy splittings in two distinct cases: at $\phi = 0$, it causes splittings between the excited states, Fig.~\ref{Fig5}(b), corresponding to the effective 2-site sweet spot, while at $\phi = \pi$, it results in splittings between the ground and excited states, Fig.~\ref{Fig5}(f), corresponding to the delocalized sweet spot.

We next analyze the microscopic model, Eqs.~(\ref{QDsH}-\ref{TunnelH}). We study the three sweet spots highlighted in Fig.~\ref{Fig3}, which feature high MP in the outermost QDs. We show the conductance as a function of the phase and voltage bias for the different types of sweet spots in Figs.~\ref{Fig6}(a, c, e). The results are qualitatively similar to those for the spinless model shown in Figs.~
\ref{Fig5}(b, d, f). The genuine 3-site sweet spot, Fig.~\ref{Fig6}(c), features degenerate excited states at $\phi = 0$ and $\phi = \pi$, unlike the other two cases. Differently from the Kitaev model, the MBSs always exhibit a finite weight on the central QD, resulting in a slight breaking of the four-fold ground-state degeneracy at $\phi = \pi$, see Figs.\ref{Fig6}(b, d). In Figs.~\ref{Fig5}(f) and \ref{Fig6}(a, c, e), we note a fainter conductance signature for the excited states at either positive or negative values of $V_L$. This is associated with the sign of the detunings that cause the delocalization of the MBS wavefunctions. In Fig.~\ref{Fig6}(a), the detuning is positive, resembling the behavior observed in Fig.~\ref{Fig5}(f), whereas Figs.~\ref{Fig6}(c, e) indicate a negative sign.

\subsection{Microwave absorption}
\label{MWSection}

We finish our analysis by studying the microwave response of the 3-site chain, based on the cavity coupling sketched in Fig.~\ref{Fig1}. This quantity, which measures transition energies between states with the same fermion parity~\cite{vayrynen2015microwave}, provides complementary information to the conductance analyzed previously. Here we focus on the odd-parity sector, which becomes the ground state for $0 <\phi < 2\pi$.

The emergence of spectral holes -- vanishing microwave absorption lines -- serves as an indicator of sweet spots \footnote{We have verified that the spectral holes vanish as the energy splits or the MP diminishes. However, this is not an exhaustive investigation of non-sweet spots.}. Here, we illustrate this for the three types of sweet spots. For the genuine 3-site sweet spot, the second and third excited states in the odd-parity subspace become degenerate at $\phi=\pi$, Fig.~\ref{Fig6}(d). This degeneracy arises from the doubly degenerate fermionic states shown in Fig.~\ref{Fig6}(c), which enable transitions involving two electrons. These degeneracies can be lifted by detuning the outermost QDs or breaking the symmetry between the left and right sides, for example, by changing the ECT and CAR values on one side. In the odd-parity sector near $\phi = \pi$, interference between the transition amplitudes to the two degenerate states results in one transition ($0 \to 3$) being amplified while the other ($0 \to 2$) is completely suppressed (although this is somewhat difficult to see in the plot because of the energy degeneracy between the second and third excited states, see Fig.~\ref{Fig9}). For the other sweet spots, there is no exact degeneracy between the excited levels, but the spectral hole is still visible near $\phi = \pi$, Figs.~\ref{Fig6}(b, f).

The three types of sweet spots can be distinguished by comparing their microwave responses. Similar to our previous analysis, the genuine 3-site sweet spot presents degeneracies between the first and second excited states at $\phi = 0$, and the second and third excited states at $\phi = \pi$, as shown in Fig.~\ref{Fig6}(d). These level crossings turn into avoided crossings for the other 2 cases, see Figs.~\ref{Fig6}(b, f). We also note that in all cases, the transition to the first excited state pronouncedly increases near $\phi = \pi$ as the excitation gap diminishes. For the effective 2-site, we observe additional spectral holes in the $0 \to 1$ transition.

\section{Conclusions} \label{conclusions}

In this work, we have analyzed the simplest extension of the minimal Kitaev chain, a chain with 3 sites, implemented in an array of 3 normal QDs connected via two QDs proximitized by superconductors. This system can feature three kinds of Majorana sweet spots with distinctive properties: an effective 2-site sweet spot; a genuine 3-site sweet spot with well-localized MBSs; and a 3-site sweet spot with delocalized MBSs. The effective 2-site regime arises when detuning the energy of the central QD in the 3-site sweet spot, creating an effective barrier between the outermost QDs. On the other hand, the sweet spot with delocalized MBSs can be achieved by detuning the outermost QDs, pushing the MBS wavefunctions toward the middle of the system.

We have shown that the conductance signatures of 3-site sweet spots differ from those for 2-site chains, allowing us to distinguish between the two cases. Finally, by considering the phase difference between the superconductors we can distinguish between all three types of sweet spots, using conductance measurements and microwave spectroscopy. We analyze these features, proposing a way to infer the Majorana localization and robustness from local measurements.

\section{Acknowledgements}
The authors acknowledge R. Aguado for fruitful discussions. R.A.D. acknowledges financial support from Conselho Nacional de Desenvolvimento Científico e Tecnológico (CNPq). M.L. acknowledges funding from the European Research Council (ERC) under the European Unions Horizon 2020 research and innovation programme under Grant Agreement No. 856526, the Swedish Research Council under Grant Agreement No. 2020-03412, and NanoLund. R.S.S acknowledges funding from the Horizon Europe Framework Program of the European Commission through the European Innovation Council Pathfinder Grant No. 101115315 (QuKiT), the Spanish Comunidad de Madrid (CM) ``Talento Program'' (Project No. 2022-T1/IND-24070), and the Spanish Ministry of Science, innovation, and Universities through Grants PID2022-140552NA-I00.

\newpage

\appendix

\section{Transport theory - Kitaev model}

The Kitaev model, Eq. (\ref{HKitaev}), disregards electron interactions. This allows us to treat the transport problem using the scattering matrix formalism. The scattering matrix, whose expression we show below, provides the probability amplitudes for the reflection and transmission of incoming electrons. 

\begin{equation}
    S = 1 - 2i \rho \pi  W^\dagger \frac{1}{E - \mathcal{H} + i \rho\pi  W W^\dagger} W,
\end{equation}
where $W$ couples the system to the leads. Considering the basis $\Phi = \begin{pmatrix}
    \Phi_{L}^e & \Phi_{R}^e & \Phi_L^h & \Phi_R^h
\end{pmatrix}^T$ for the leads, where $\Phi_\alpha^\beta$ represents the wavefunction of a particle of type $\beta = e, h$ in the $\alpha = L, R$ lead, the $W$ matrix reads

\begin{equation}
    W = \begin{pmatrix}
        t_L & 0 & 0 & 0 \\ 0 & 0 & 0 & 0\\ 0 & t_R & 0 & 0 \\ 0 & 0 & -t_L & 0 \\ 0 & 0 & 0 & 0 \\ 0 & 0 & 0 & -t_R
    \end{pmatrix},
\end{equation}
where $t_\alpha$ is the hopping between lead $\alpha$ and the system. In addition, the Bogoliubov-de Gennes Hamiltonian is

\begin{equation}
    \mathcal{H} = \begin{pmatrix}
        \epsilon_1 & \tau_{12} & 0 & 0 & \delta & 0 \\
        \tau_{12} & \epsilon_2 & \tau_{23} & -\delta & 0 & \delta e^{i \phi} \\
        0 & \tau_{23} & \epsilon_3 & 0 & -\delta e^{i \phi} & 0 \\
        0 & -\delta & 0 & -\epsilon_1 & -\tau_{12} & 0 \\
        \delta & 0 & -\delta e^{-i \phi} & -\tau_{12} & -\epsilon_2 & -\tau_{23} \\
        0 & \delta e^{-i\phi} & 0 & 0 & -\tau_{23} & -\epsilon_3 \\
    \end{pmatrix},
\end{equation}
where we defined the basis as $\psi = \begin{pmatrix}
    c_1 & c_2 & c_3 & c_1^\dagger & c_2^\dagger & c_3^\dagger
\end{pmatrix}^T$.

After obtaining the scattering matrix, it is possible to calculate the current using the expression \cite{maiani2022conductance}

\begin{equation} \label{I}
\begin{split}
		 I_\alpha &= \frac{e}{h} \int dE \left(2 A_{\alpha \alpha} + T_{\Bar{\alpha} \alpha} + A_{\Bar{\alpha} \alpha} \right) \Tilde{f}(\mu_\alpha) \\
   &- \frac{e}{h} \int dE \left( T_{\alpha \Bar{\alpha}} - A_{\alpha \Bar{\alpha}} \right) \Tilde{f}(\mu_{\Bar{\alpha}}),
\end{split}
\end{equation}
where $\alpha \neq \Bar{\alpha} = L, R$, $\Tilde{f}(\mu_\alpha) = f(E - \mu_\alpha) - f(E)$, and $f(E) = \left[1 + e^{(E/k_B T)} \right]^{-1}$ is the Fermi function, and $\mu_\alpha = e V_\alpha$, given by the applied voltage bias. The coefficients $A_{ij}$ and $T_{ij}$, given by the s-matrix, represent the transfer of an incoming electron from lead $i$ to lead $j$ as a hole or electron, respectively.

\section{microwave absorption for the Kitaev model} \label{MWAppendixKitaev}

In this Appendix, we discuss the microwave absorption spectrum for the Kitaev model. We first obtain analytical expressions using perturbation theory that corroborates the findings of Ref. \cite{pino2024minimal}, where the emergence of spectral holes in the microwave spectrum was associated with the presence of MBSs. Then, we perform numerical simulations and show the spectral holes at $\phi = \pi$ for the transitions between the ground state and the second and third excited states.

We start by projecting the Hamiltonian into the even and odd subspaces, which are uncoupled since the above Hamiltonian commutes with the parity operator. By doing that, in the basis $\psi_e = \{\ket{0, 0, 0}, \ket{1, 1, 0}, \ket{1, 0, 1}, \ket{0, 1, 1} \}$ and $\psi_e = \{\ket{1, 0, 0}, \ket{0, 1, 0}, \ket{0, 0, 1}, \ket{1, 1, 1} \}$ ($\ket{1, 1, 1} = c_1^\dagger c_2^\dagger c_3^\dagger \ket{0, 0, 0}$), we obtain

\begin{equation}
    \mathcal{H} = \begin{pmatrix}
        \mathcal{H}_e & 0 \\ 0 & \mathcal{H}_o
    \end{pmatrix},
\end{equation}
where

\begin{equation}
    \mathcal{H}_e = \begin{pmatrix}
        0 & \delta & 0 & \delta e^{-i \phi} \\ \delta & \Lambda_{12} & \tau & 0 \\ 0 & \tau & \Lambda_{13} & \tau \\ \delta e^{i \phi} & 0 & \tau & \Lambda_{23} 
        \end{pmatrix},
\end{equation}
\begin{equation}
\mathcal{H}_o = \begin{pmatrix}
            \epsilon_1 & \tau & 0 & \delta e^{-i \phi} \\ \tau & \epsilon_2 & \tau & 0 \\ 0 & \tau & \epsilon_3 &\delta \\ \delta e^{i \phi} & 0 & \delta & \Lambda_{123}
    \end{pmatrix},
\end{equation}
and $\Lambda_{ij(k)} = \epsilon_i + \epsilon_j (+ \epsilon_k)$.

Initially, let us consider the sweet spot, i.e., $\tau = 1$, all $\epsilon_i = 0$ ($\delta = 1$ throughout). The Hamiltonian then becomes

\begin{equation}
    \mathcal{H}_e^{(0)} =  \mathcal{H}_o^{(0)}=  \begin{pmatrix}
        0 & 1 & 0 & e^{-i \phi} \\ 1 & 0 & 1 & 0 \\ 0 & 1 & 0 & 1 \\  e^{i \phi} & 0 & 1 & 0 
        \end{pmatrix}.
\end{equation}

The eigenenergies are $E_i = \pm \sqrt{2 \pm \sqrt{2(1 + \cos(\phi))} } = \pm \sqrt{2(1 \pm |\cos(\phi/2)|)}$, and the eigenvectors are

\begin{equation}
    \psi_i = \frac{c}{E_i^2 -1 + e^{-i \phi}}\begin{pmatrix}
        1 + (E_i^2 - 1) e^{- i\phi} \\ E_i(1 + e^{-i \phi}) \\ E_i^2 -1 + e^{-i \phi} \\ E_i(E_i^2 - 2).
    \end{pmatrix}
\end{equation}

Considering $\phi \leq \pi$, $|\cos(\phi/2)| = \cos(\phi/2)$. The eigenenergies become, $E_0 = - \cos(\phi/4)$, $E_1 = - \sin(\phi/4)$, $E_2 =  \sin(\phi/4)$, and $E_3 = \cos(\phi/4)$. The eigenvectors are

\begin{equation}
\begin{split}
    \psi_0 &= \frac{1}{2}\begin{pmatrix}
        e^{-i \frac{\phi}{2}} \\ - e^{-i \frac{\phi}{4}} \\ 1 \\ -e^{i \frac{\phi}{4}},
    \end{pmatrix}, \,\,     \psi_1 = \frac{1}{2}\begin{pmatrix}
        -e^{-i \frac{\phi}{2}} \\ -i e^{-i \frac{\phi}{4}} \\ 1 \\ i e^{i \frac{\phi}{4}}
    \end{pmatrix}, \\ 
    &\psi_2 = \frac{1}{2}\begin{pmatrix}
        -e^{-i \frac{\phi}{2}} \\ i e^{-i \frac{\phi}{4}} \\ 1 \\ -i e^{i \frac{\phi}{4}}   \end{pmatrix}, \,\,     \psi_3 = \frac{1}{2}\begin{pmatrix}
        e^{-i \frac{\phi}{2}} \\  e^{-i \frac{\phi}{4}} \\ 1 \\ e^{i \frac{\phi}{4}}
    \end{pmatrix}.
\end{split}
\end{equation}

We are interested in the quantity $t_{02} = \psi_2^\dagger (-i \partial_\phi) \psi_0$. In this case, the result yields

\begin{equation}
    t_{02} = \frac{1-i}{8}.
\end{equation}

To appreciate the phase dependency on the above transition, we move the system from the sweet spot by considering a detuning of the site energies $\epsilon_i = \epsilon$. Additionally, we set $\phi = \pi$ and consider small phase deviations as a perturbation, such that $\phi - \pi \ll 1$. In this case, the even-parity Hamiltonian becomes 

\begin{equation}
    \mathcal{H}_e = \mathcal{H}_{e,0} + V,
\end{equation}
\begin{equation}
\mathcal{H}_{e,0} =
    \begin{pmatrix}
        0 & 1 & 0 & -1 \\ 1 & 2\epsilon & 1 & 0 \\ 0 & 1 & 2\epsilon & 1 \\ -1 & 0 & 1 & 2\epsilon 
        \end{pmatrix} ,
\end{equation}
\begin{equation}
    V =\begin{pmatrix}
        0 & 0 & 0 & 1-e^{-i (\phi - \pi)} \\ 0 & 0 & 0 & 0 \\ 0 & 0 & 0 & 0 \\ 1-e^{i(\phi - \pi)} & 0 & 0 & 0 
        \end{pmatrix}.
\end{equation}
The eigenvalues and eigenvectors of $\mathcal{H}_{e,0}$ are

\begin{equation}
    E_0 = 2\epsilon -\sqrt{2}, \,\,\, \psi_0 = \frac{1}{2}\begin{pmatrix}
        0 & 1 & -\sqrt{2} & 1
    \end{pmatrix}^T,
\end{equation}
\begin{equation}
    E_1 = \epsilon -\sqrt{2 + \epsilon^2}, \,\,\, \psi_1 = \frac{1}{\sqrt{E_3^2 + 2}}\begin{pmatrix}
        E_3 & -1 & 0 & 1
    \end{pmatrix}^T,
\end{equation}
\begin{equation}
    E_2 = 2\epsilon +\sqrt{2}, \,\,\, \psi_2 = \frac{1}{2}\begin{pmatrix}
        0 & 1 & \sqrt{2} & 1
    \end{pmatrix}^T,
\end{equation}
\begin{equation}
    E_3 = \epsilon -\sqrt{2 + \epsilon^2}, \,\,\, \psi_3 = \frac{1}{\sqrt{E_1^2 + 2}}\begin{pmatrix}
        E_1 & -1 & 0 & 1
    \end{pmatrix}^T.
\end{equation}

The wavefunctions $\psi_0$ and $\psi_2$ after the first order correction become

\begin{equation}
    \Psi_0 = \psi_0 + (a_+ \psi_1 + a_- \psi_3) \left(1 - e^{-i(\phi - \pi)}\right),
\end{equation}
\begin{equation}
    \Psi_2 = \psi_2 + (b_- \psi_1 + b_+ \psi_3)\left(1 - e^{-i(\phi - \pi)}\right),
\end{equation}
where

\begin{equation}
    a_+ = \frac{E_3}{2(E_0 - E_1)\sqrt{E_3^2 + 2}}, \,\,\, a_- = \frac{E_1}{2(E_0 - E_3)\sqrt{E_1^2 + 2}},
\end{equation}
\begin{equation}
    b_- = \frac{E_3}{2(E_2 - E_1)\sqrt{E_3^2 + 2}}, \,\,\, b_+ = \frac{E_1}{2(E_2 - E_3)\sqrt{E_1^2 + 2}}.
\end{equation}
Then we calculate the $t_{02} = -i \Psi_2^\dagger \partial_\phi \Psi_0$ transition again, obtaining
\begin{equation} \label{t02 perturbation}
     t_{02} = (b_- a_+ + b_+ a_-)[\cos(\phi - \pi) - 1 -i \sin(\phi - \pi)].
\end{equation}
This result shows that for $\tau = 1$, the Kitaev model presents a spectral hole in $t_{02}$ at $\phi = \pi$, similar to the results obtained for sweet spots in the microscopic model, presented in Fig.~\ref{Fig6}. The above expression describes this transition within the odd (even) parity sector for $\epsilon < 0$ ($\epsilon > 0$), depending on which one has the lowest energy at $\phi = \pi$. 

\begin{figure}
\centering
\includegraphics[width=1\linewidth]{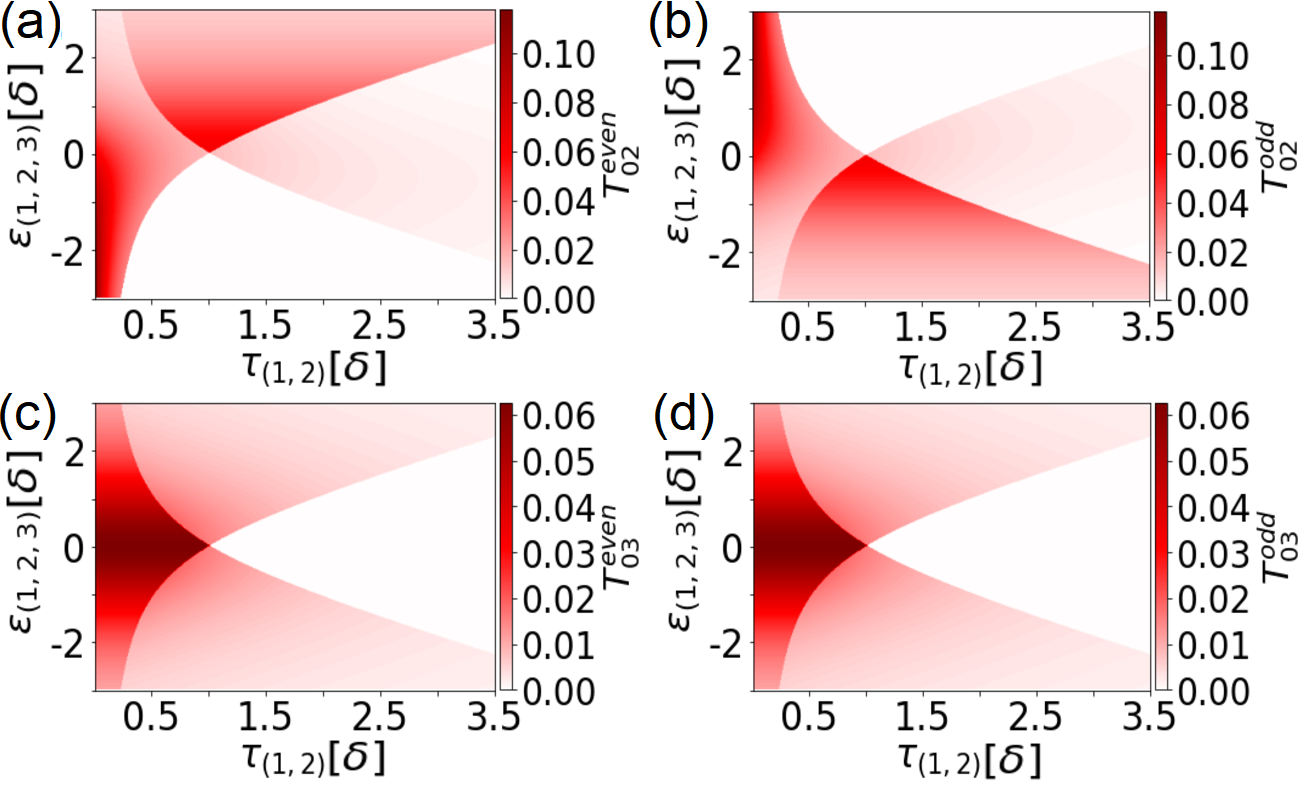}
\caption{Transition amplitudes for the 3-sites Kitaev chain as a function of the on-site energies ($\epsilon$) and hopping ($\tau$).}
\label{transitions 2 and 3}
\end{figure}

We now generalize the results by plotting the transitions from the ground state to the second and third excited states, $T_{0i} = |t_{0i}|^2$, at $\phi = \pi$ as we vary $\tau$ and $\epsilon$ in Fig.~\ref{transitions 2 and 3}. We note that depending on the values for the parameters, it can be either the transition to the second or third excited states that vanish at $\phi = \pi$. For instance, for larger values of $\tau$ and small values of $\epsilon$, see Figs.~\ref{transitions 2 and 3}(c-d), it is $T_{03}$ that vanishes, which corroborates the spectral hole observed in some of the sweet spots considered in the microscopic model, for instance, the results for the delocalized sweet spot in the third row of Fig.~\ref{Fig6}. From the two figures, we conclude that the $0 \to 3$ transition presents a spectral hole at $\phi = \pi$ in cases where the MBSs present delocalized wave functions, which in the Kitaev model occurs for larger values of $\tau/\epsilon$, see the white regions in Figs.~\ref{transitions 2 and 3}(c-d). 

\section{Microwave transitions for the microscopic model}

In this section, we present additional plots to complement the phase dependence of the results shown in Fig.~\ref{Fig6} for the microscopic model.

\begin{figure}[hbt!]
\centering \includegraphics[width=\linewidth]{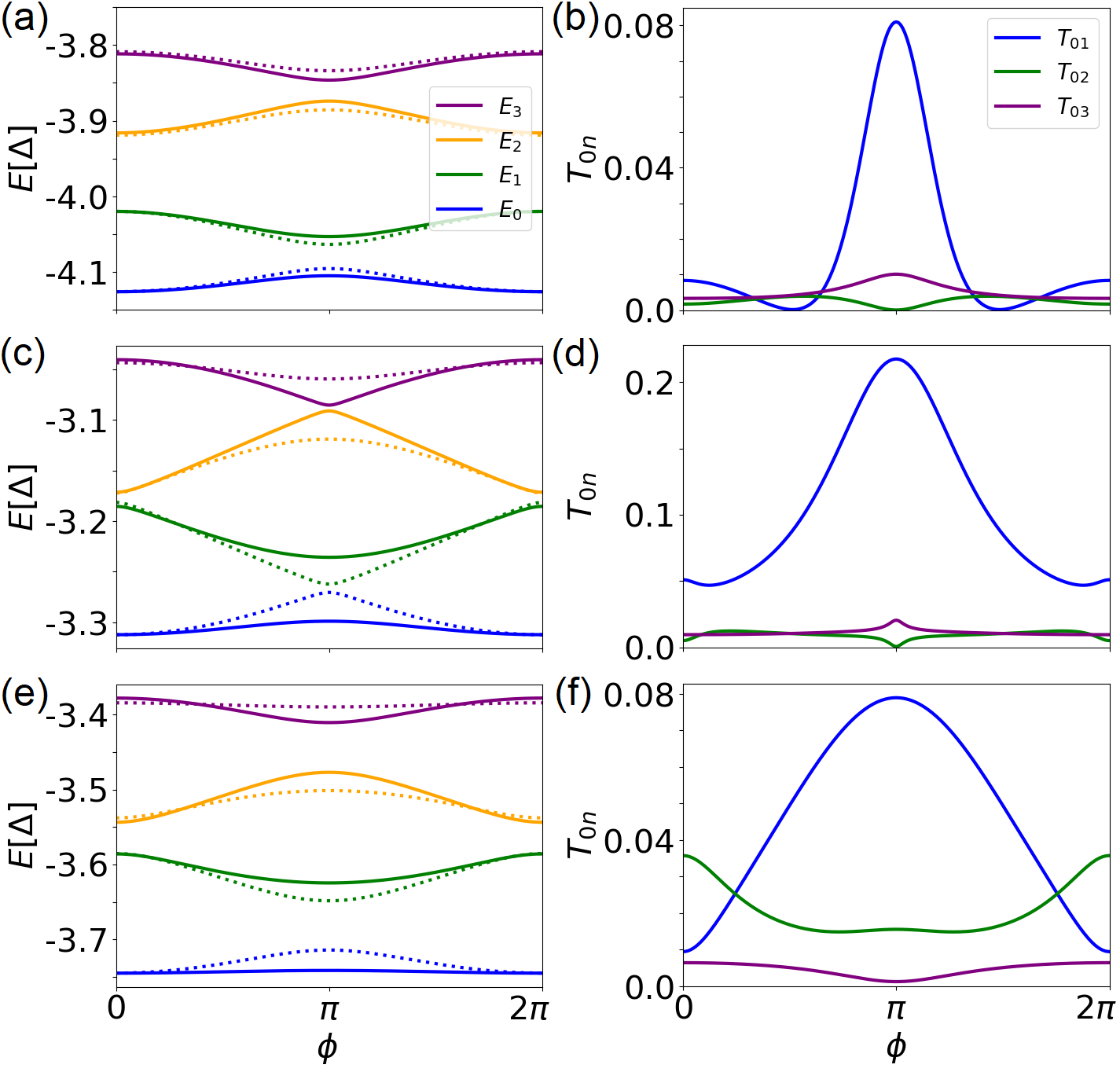}
\caption{(a), (c), and (e) energies for the odd (solid lines) and even (dotted lines) parity sectors as a function of $\phi$. (b), (d), and (f) Transition amplitudes, $T_{0n} = |\bra{n}\partial_\phi \ket{0}|^2$, for the lowest-energy (odd) parity sector as a function of $\phi$. The first to third rows show the results for the green, purple, and cyan sweet spots depicted in Fig.~\ref{Fig3}, respectively.}
\label{Fig9}
\end{figure}

The first column of Fig.~\ref{Fig9} displays the energies for the odd (solid lines) and even (dotted lines) parity sectors. In all cases, the lowest-energy mode of the odd parity sector becomes the ground state. Consequently, we focus on this parity sector when calculating the transition amplitudes $T_{0n}$, presented in the second column. The first, second, and third rows correspond to the effective 2-site sweet spot (green), genuine 3-site sweet spot (purple), and delocalized sweet spot (cyan) sweet spots, respectively. In all cases, the transition to the second or third excited states vanishes near $\phi = \pi$.

\bibliography{bibliography.bib}

\providecommand{\noopsort}[1]{}\providecommand{\singleletter}[1]{#1}%
\begin{thebibliography}{53}%
\makeatletter
\providecommand \@ifxundefined [1]{%
 \@ifx{#1\undefined}
}%
\providecommand \@ifnum [1]{%
 \ifnum #1\expandafter \@firstoftwo
 \else \expandafter \@secondoftwo
 \fi
}%
\providecommand \@ifx [1]{%
 \ifx #1\expandafter \@firstoftwo
 \else \expandafter \@secondoftwo
 \fi
}%
\providecommand \natexlab [1]{#1}%
\providecommand \enquote  [1]{``#1''}%
\providecommand \bibnamefont  [1]{#1}%
\providecommand \bibfnamefont [1]{#1}%
\providecommand \citenamefont [1]{#1}%
\providecommand \href@noop [0]{\@secondoftwo}%
\providecommand \href [0]{\begingroup \@sanitize@url \@href}%
\providecommand \@href[1]{\@@startlink{#1}\@@href}%
\providecommand \@@href[1]{\endgroup#1\@@endlink}%
\providecommand \@sanitize@url [0]{\catcode `\\12\catcode `\$12\catcode `\&12\catcode `\#12\catcode `\^12\catcode `\_12\catcode `\%12\relax}%
\providecommand \@@startlink[1]{}%
\providecommand \@@endlink[0]{}%
\providecommand \url  [0]{\begingroup\@sanitize@url \@url }%
\providecommand \@url [1]{\endgroup\@href {#1}{\urlprefix }}%
\providecommand \urlprefix  [0]{URL }%
\providecommand \Eprint [0]{\href }%
\providecommand \doibase [0]{https://doi.org/}%
\providecommand \selectlanguage [0]{\@gobble}%
\providecommand \bibinfo  [0]{\@secondoftwo}%
\providecommand \bibfield  [0]{\@secondoftwo}%
\providecommand \translation [1]{[#1]}%
\providecommand \BibitemOpen [0]{}%
\providecommand \bibitemStop [0]{}%
\providecommand \bibitemNoStop [0]{.\EOS\space}%
\providecommand \EOS [0]{\spacefactor3000\relax}%
\providecommand \BibitemShut  [1]{\csname bibitem#1\endcsname}%
\let\auto@bib@innerbib\@empty
\bibitem [{\citenamefont {Leijnse}\ and\ \citenamefont {Flensberg}(2012{\natexlab{a}})}]{leijnse2012introduction}%
  \BibitemOpen
  \bibfield  {author} {\bibinfo {author} {\bibfnamefont {M.}~\bibnamefont {Leijnse}}\ and\ \bibinfo {author} {\bibfnamefont {K.}~\bibnamefont {Flensberg}},\ }\bibfield  {title} {\bibinfo {title} {Introduction to topological superconductivity and {M}ajorana fermions},\ }\href {https://dx.doi.org/10.1088/0268-1242/27/12/124003} {\bibfield  {journal} {\bibinfo  {journal} {Semiconductor Science and Technology}\ }\textbf {\bibinfo {volume} {27}},\ \bibinfo {pages} {124003} (\bibinfo {year} {2012}{\natexlab{a}})}\BibitemShut {NoStop}%
\bibitem [{\citenamefont {Alicea}(2012)}]{alicea2012new}%
  \BibitemOpen
  \bibfield  {author} {\bibinfo {author} {\bibfnamefont {J.}~\bibnamefont {Alicea}},\ }\bibfield  {title} {\bibinfo {title} {New directions in the pursuit of {M}ajorana fermions in solid state systems},\ }\href {https://dx.doi.org/10.1088/0034-4885/75/7/076501} {\bibfield  {journal} {\bibinfo  {journal} {Reports on progress in physics}\ }\textbf {\bibinfo {volume} {75}},\ \bibinfo {pages} {076501} (\bibinfo {year} {2012})}\BibitemShut {NoStop}%
\bibitem [{\citenamefont {Aguado}(2017)}]{aguado2017}%
  \BibitemOpen
  \bibfield  {author} {\bibinfo {author} {\bibfnamefont {R.}~\bibnamefont {Aguado}},\ }\bibfield  {title} {\bibinfo {title} {Majorana quasiparticles in condensed matter},\ }\href {https://doi.org/10.1393/ncr/i2017-10141-9} {\bibfield  {journal} {\bibinfo  {journal} {La Rivista del Nuovo Cimento}\ }\textbf {\bibinfo {volume} {40}},\ \bibinfo {pages} {523} (\bibinfo {year} {2017})}\BibitemShut {NoStop}%
\bibitem [{\citenamefont {Beenakker}(2020)}]{beenakker2020search}%
  \BibitemOpen
  \bibfield  {author} {\bibinfo {author} {\bibfnamefont {C.}~\bibnamefont {Beenakker}},\ }\bibfield  {title} {\bibinfo {title} {Search for non-{A}belian {M}ajorana braiding statistics in superconductors},\ }\href@noop {} {\bibfield  {journal} {\bibinfo  {journal} {SciPost Physics Lecture Notes}\ ,\ \bibinfo {pages} {015}} (\bibinfo {year} {2020})}\BibitemShut {NoStop}%
\bibitem [{\citenamefont {Prada}\ \emph {et~al.}(2020)\citenamefont {Prada}, \citenamefont {San-Jose}, \citenamefont {de~Moor}, \citenamefont {Geresdi}, \citenamefont {Lee}, \citenamefont {Klinovaja}, \citenamefont {Loss}, \citenamefont {Nyg{\aa}rd}, \citenamefont {Aguado},\ and\ \citenamefont {Kouwenhoven}}]{prada2020andreev}%
  \BibitemOpen
  \bibfield  {author} {\bibinfo {author} {\bibfnamefont {E.}~\bibnamefont {Prada}}, \bibinfo {author} {\bibfnamefont {P.}~\bibnamefont {San-Jose}}, \bibinfo {author} {\bibfnamefont {M.~W.}\ \bibnamefont {de~Moor}}, \bibinfo {author} {\bibfnamefont {A.}~\bibnamefont {Geresdi}}, \bibinfo {author} {\bibfnamefont {E.~J.}\ \bibnamefont {Lee}}, \bibinfo {author} {\bibfnamefont {J.}~\bibnamefont {Klinovaja}}, \bibinfo {author} {\bibfnamefont {D.}~\bibnamefont {Loss}}, \bibinfo {author} {\bibfnamefont {J.}~\bibnamefont {Nyg{\aa}rd}}, \bibinfo {author} {\bibfnamefont {R.}~\bibnamefont {Aguado}},\ and\ \bibinfo {author} {\bibfnamefont {L.~P.}\ \bibnamefont {Kouwenhoven}},\ }\bibfield  {title} {\bibinfo {title} {From andreev to majorana bound states in hybrid superconductor--semiconductor nanowires},\ }\href@noop {} {\bibfield  {journal} {\bibinfo  {journal} {Nature Reviews Physics}\ }\textbf {\bibinfo {volume} {2}},\ \bibinfo {pages} {575} (\bibinfo {year} {2020})}\BibitemShut {NoStop}%
\bibitem [{\citenamefont {Flensberg}\ \emph {et~al.}(2021)\citenamefont {Flensberg}, \citenamefont {von Oppen},\ and\ \citenamefont {Stern}}]{flensberg2021engineered}%
  \BibitemOpen
  \bibfield  {author} {\bibinfo {author} {\bibfnamefont {K.}~\bibnamefont {Flensberg}}, \bibinfo {author} {\bibfnamefont {F.}~\bibnamefont {von Oppen}},\ and\ \bibinfo {author} {\bibfnamefont {A.}~\bibnamefont {Stern}},\ }\bibfield  {title} {\bibinfo {title} {Engineered platforms for topological superconductivity and {M}ajorana zero modes},\ }\href {https://doi.org/10.1038/s41578-021-00336-6} {\bibfield  {journal} {\bibinfo  {journal} {Nature Reviews Materials}\ }\textbf {\bibinfo {volume} {6}},\ \bibinfo {pages} {944} (\bibinfo {year} {2021})}\BibitemShut {NoStop}%
\bibitem [{\citenamefont {Das~Sarma}(2023)}]{das2023search}%
  \BibitemOpen
  \bibfield  {author} {\bibinfo {author} {\bibfnamefont {S.}~\bibnamefont {Das~Sarma}},\ }\bibfield  {title} {\bibinfo {title} {In search of {M}ajorana},\ }\href {https://doi.org/10.1038/s41567-022-01900-9} {\bibfield  {journal} {\bibinfo  {journal} {Nature Physics}\ }\textbf {\bibinfo {volume} {19}},\ \bibinfo {pages} {165} (\bibinfo {year} {2023})}\BibitemShut {NoStop}%
\bibitem [{\citenamefont {Kouwenhoven}(0)}]{kouwenhoven2024perspective}%
  \BibitemOpen
  \bibfield  {author} {\bibinfo {author} {\bibfnamefont {L.}~\bibnamefont {Kouwenhoven}},\ }\bibfield  {title} {\bibinfo {title} {Perspective on {M}ajorana bound-states in hybrid superconductor-semiconductor nanowires},\ }\href {https://doi.org/10.1142/S0217984925400020} {\bibfield  {journal} {\bibinfo  {journal} {Modern Physics Letters B}\ }\textbf {\bibinfo {volume} {0}},\ \bibinfo {pages} {2540002} (\bibinfo {year} {0})}\BibitemShut {NoStop}%
\bibitem [{\citenamefont {Kitaev}(2001)}]{Kitaev2001}%
  \BibitemOpen
  \bibfield  {author} {\bibinfo {author} {\bibfnamefont {A.~Y.}\ \bibnamefont {Kitaev}},\ }\bibfield  {title} {\bibinfo {title} {Unpaired {M}ajorana fermions in quantum wires},\ }\href {https://doi.org/10.1070/1063-7869/44/10S/S29} {\bibfield  {journal} {\bibinfo  {journal} {Physics-Uspekhi}\ }\textbf {\bibinfo {volume} {44}},\ \bibinfo {pages} {131} (\bibinfo {year} {2001})}\BibitemShut {NoStop}%
\bibitem [{\citenamefont {Leijnse}\ and\ \citenamefont {Flensberg}(2012{\natexlab{b}})}]{Leijnse2012}%
  \BibitemOpen
  \bibfield  {author} {\bibinfo {author} {\bibfnamefont {M.}~\bibnamefont {Leijnse}}\ and\ \bibinfo {author} {\bibfnamefont {K.}~\bibnamefont {Flensberg}},\ }\bibfield  {title} {\bibinfo {title} {Parity qubits and poor man's {M}ajorana bound states in double quantum dots},\ }\href {https://doi.org/10.1103/PhysRevB.86.134528} {\bibfield  {journal} {\bibinfo  {journal} {Phys. Rev. B}\ }\textbf {\bibinfo {volume} {86}},\ \bibinfo {pages} {134528} (\bibinfo {year} {2012}{\natexlab{b}})}\BibitemShut {NoStop}%
\bibitem [{\citenamefont {Sau}\ and\ \citenamefont {Sarma}(2012)}]{Sau2012}%
  \BibitemOpen
  \bibfield  {author} {\bibinfo {author} {\bibfnamefont {J.~D.}\ \bibnamefont {Sau}}\ and\ \bibinfo {author} {\bibfnamefont {S.~D.}\ \bibnamefont {Sarma}},\ }\bibfield  {title} {\bibinfo {title} {Realizing a robust practical {M}ajorana chain in a quantum-dot-superconductor linear array},\ }\href {https://doi.org/10.1038/ncomms1966} {\bibfield  {journal} {\bibinfo  {journal} {Nature communications}\ }\textbf {\bibinfo {volume} {3}},\ \bibinfo {pages} {964} (\bibinfo {year} {2012})}\BibitemShut {NoStop}%
\bibitem [{\citenamefont {Fulga}\ \emph {et~al.}(2013)\citenamefont {Fulga}, \citenamefont {Haim}, \citenamefont {Akhmerov},\ and\ \citenamefont {Oreg}}]{Fulga2013}%
  \BibitemOpen
  \bibfield  {author} {\bibinfo {author} {\bibfnamefont {I.~C.}\ \bibnamefont {Fulga}}, \bibinfo {author} {\bibfnamefont {A.}~\bibnamefont {Haim}}, \bibinfo {author} {\bibfnamefont {A.~R.}\ \bibnamefont {Akhmerov}},\ and\ \bibinfo {author} {\bibfnamefont {Y.}~\bibnamefont {Oreg}},\ }\bibfield  {title} {\bibinfo {title} {Adaptive tuning of {M}ajorana fermions in a quantum dot chain},\ }\href {https://doi.org/10.1088/1367-2630/15/4/045020} {\bibfield  {journal} {\bibinfo  {journal} {New journal of physics}\ }\textbf {\bibinfo {volume} {15}},\ \bibinfo {pages} {045020} (\bibinfo {year} {2013})}\BibitemShut {NoStop}%
\bibitem [{\citenamefont {Liu}\ \emph {et~al.}(2022)\citenamefont {Liu}, \citenamefont {Wang}, \citenamefont {Dvir},\ and\ \citenamefont {Wimmer}}]{liu2022tunable}%
  \BibitemOpen
  \bibfield  {author} {\bibinfo {author} {\bibfnamefont {C.-X.}\ \bibnamefont {Liu}}, \bibinfo {author} {\bibfnamefont {G.}~\bibnamefont {Wang}}, \bibinfo {author} {\bibfnamefont {T.}~\bibnamefont {Dvir}},\ and\ \bibinfo {author} {\bibfnamefont {M.}~\bibnamefont {Wimmer}},\ }\bibfield  {title} {\bibinfo {title} {Tunable superconducting coupling of quantum dots via andreev bound states in semiconductor-superconductor nanowires},\ }\href {https://doi.org/10.1103/PhysRevLett.129.267701} {\bibfield  {journal} {\bibinfo  {journal} {Phys. Rev. Lett.}\ }\textbf {\bibinfo {volume} {129}},\ \bibinfo {pages} {267701} (\bibinfo {year} {2022})}\BibitemShut {NoStop}%
\bibitem [{\citenamefont {Tsintzis}\ \emph {et~al.}(2022)\citenamefont {Tsintzis}, \citenamefont {Souto},\ and\ \citenamefont {Leijnse}}]{tsintzis2022creating}%
  \BibitemOpen
  \bibfield  {author} {\bibinfo {author} {\bibfnamefont {A.}~\bibnamefont {Tsintzis}}, \bibinfo {author} {\bibfnamefont {R.~S.}\ \bibnamefont {Souto}},\ and\ \bibinfo {author} {\bibfnamefont {M.}~\bibnamefont {Leijnse}},\ }\bibfield  {title} {\bibinfo {title} {Creating and detecting poor man's {M}ajorana bound states in interacting quantum dots},\ }\href {https://link.aps.org/doi/10.1103/PhysRevB.106.L201404} {\bibfield  {journal} {\bibinfo  {journal} {Physical Review B}\ }\textbf {\bibinfo {volume} {106}},\ \bibinfo {pages} {L201404} (\bibinfo {year} {2022})}\BibitemShut {NoStop}%
\bibitem [{\citenamefont {Liu}\ \emph {et~al.}(2024)\citenamefont {Liu}, \citenamefont {Bozkurt}, \citenamefont {Zatelli}, \citenamefont {ten Haaf}, \citenamefont {Dvir},\ and\ \citenamefont {Wimmer}}]{liu2024enhancing}%
  \BibitemOpen
  \bibfield  {author} {\bibinfo {author} {\bibfnamefont {C.-X.}\ \bibnamefont {Liu}}, \bibinfo {author} {\bibfnamefont {A.~M.}\ \bibnamefont {Bozkurt}}, \bibinfo {author} {\bibfnamefont {F.}~\bibnamefont {Zatelli}}, \bibinfo {author} {\bibfnamefont {S.~L.}\ \bibnamefont {ten Haaf}}, \bibinfo {author} {\bibfnamefont {T.}~\bibnamefont {Dvir}},\ and\ \bibinfo {author} {\bibfnamefont {M.}~\bibnamefont {Wimmer}},\ }\bibfield  {title} {\bibinfo {title} {Enhancing the excitation gap of a quantum-dot-based kitaev chain},\ }\href {https://doi.org/10.1038/s42005-024-01715-5} {\bibfield  {journal} {\bibinfo  {journal} {Communications Physics}\ }\textbf {\bibinfo {volume} {7}},\ \bibinfo {pages} {235} (\bibinfo {year} {2024})}\BibitemShut {NoStop}%
\bibitem [{\citenamefont {Dvir}\ \emph {et~al.}(2023)\citenamefont {Dvir}, \citenamefont {Wang}, \citenamefont {van Loo}, \citenamefont {Liu}, \citenamefont {Mazur}, \citenamefont {Bordin}, \citenamefont {Ten~Haaf}, \citenamefont {Wang}, \citenamefont {van Driel}, \citenamefont {Zatelli} \emph {et~al.}}]{Dvir2023}%
  \BibitemOpen
  \bibfield  {author} {\bibinfo {author} {\bibfnamefont {T.}~\bibnamefont {Dvir}}, \bibinfo {author} {\bibfnamefont {G.}~\bibnamefont {Wang}}, \bibinfo {author} {\bibfnamefont {N.}~\bibnamefont {van Loo}}, \bibinfo {author} {\bibfnamefont {C.-X.}\ \bibnamefont {Liu}}, \bibinfo {author} {\bibfnamefont {G.~P.}\ \bibnamefont {Mazur}}, \bibinfo {author} {\bibfnamefont {A.}~\bibnamefont {Bordin}}, \bibinfo {author} {\bibfnamefont {S.~L.}\ \bibnamefont {Ten~Haaf}}, \bibinfo {author} {\bibfnamefont {J.-Y.}\ \bibnamefont {Wang}}, \bibinfo {author} {\bibfnamefont {D.}~\bibnamefont {van Driel}}, \bibinfo {author} {\bibfnamefont {F.}~\bibnamefont {Zatelli}}, \emph {et~al.},\ }\bibfield  {title} {\bibinfo {title} {Realization of a minimal {K}itaev chain in coupled quantum dots},\ }\href {https://doi.org/10.1038/s41586-022-05585-1} {\bibfield  {journal} {\bibinfo  {journal} {Nature}\ }\textbf {\bibinfo {volume} {614}},\ \bibinfo {pages} {445} (\bibinfo {year} {2023})}\BibitemShut {NoStop}%
\bibitem [{\citenamefont {Bordin}\ \emph {et~al.}(2023)\citenamefont {Bordin}, \citenamefont {Wang}, \citenamefont {Liu}, \citenamefont {ten Haaf}, \citenamefont {van Loo}, \citenamefont {Mazur}, \citenamefont {Xu}, \citenamefont {van Driel}, \citenamefont {Zatelli}, \citenamefont {Gazibegovic}, \citenamefont {Badawy}, \citenamefont {Bakkers}, \citenamefont {Wimmer}, \citenamefont {Kouwenhoven},\ and\ \citenamefont {Dvir}}]{bordin2023tunable}%
  \BibitemOpen
  \bibfield  {author} {\bibinfo {author} {\bibfnamefont {A.}~\bibnamefont {Bordin}}, \bibinfo {author} {\bibfnamefont {G.}~\bibnamefont {Wang}}, \bibinfo {author} {\bibfnamefont {C.-X.}\ \bibnamefont {Liu}}, \bibinfo {author} {\bibfnamefont {S.~L.~D.}\ \bibnamefont {ten Haaf}}, \bibinfo {author} {\bibfnamefont {N.}~\bibnamefont {van Loo}}, \bibinfo {author} {\bibfnamefont {G.~P.}\ \bibnamefont {Mazur}}, \bibinfo {author} {\bibfnamefont {D.}~\bibnamefont {Xu}}, \bibinfo {author} {\bibfnamefont {D.}~\bibnamefont {van Driel}}, \bibinfo {author} {\bibfnamefont {F.}~\bibnamefont {Zatelli}}, \bibinfo {author} {\bibfnamefont {S.}~\bibnamefont {Gazibegovic}}, \bibinfo {author} {\bibfnamefont {G.}~\bibnamefont {Badawy}}, \bibinfo {author} {\bibfnamefont {E.~P. A.~M.}\ \bibnamefont {Bakkers}}, \bibinfo {author} {\bibfnamefont {M.}~\bibnamefont {Wimmer}}, \bibinfo {author} {\bibfnamefont {L.~P.}\ \bibnamefont {Kouwenhoven}},\ and\ \bibinfo {author} {\bibfnamefont {T.}~\bibnamefont {Dvir}},\ }\bibfield  {title} {\bibinfo
  {title} {Tunable crossed andreev reflection and elastic cotunneling in hybrid nanowires},\ }\href {https://doi.org/10.1103/PhysRevX.13.031031} {\bibfield  {journal} {\bibinfo  {journal} {Phys. Rev. X}\ }\textbf {\bibinfo {volume} {13}},\ \bibinfo {pages} {031031} (\bibinfo {year} {2023})}\BibitemShut {NoStop}%
\bibitem [{\citenamefont {Bordin}\ \emph {et~al.}(2024{\natexlab{a}})\citenamefont {Bordin}, \citenamefont {Li}, \citenamefont {van Driel}, \citenamefont {Wolff}, \citenamefont {Wang}, \citenamefont {ten Haaf}, \citenamefont {Wang}, \citenamefont {van Loo}, \citenamefont {Kouwenhoven},\ and\ \citenamefont {Dvir}}]{bordin2024crossed}%
  \BibitemOpen
  \bibfield  {author} {\bibinfo {author} {\bibfnamefont {A.}~\bibnamefont {Bordin}}, \bibinfo {author} {\bibfnamefont {X.}~\bibnamefont {Li}}, \bibinfo {author} {\bibfnamefont {D.}~\bibnamefont {van Driel}}, \bibinfo {author} {\bibfnamefont {J.~C.}\ \bibnamefont {Wolff}}, \bibinfo {author} {\bibfnamefont {Q.}~\bibnamefont {Wang}}, \bibinfo {author} {\bibfnamefont {S.~L.~D.}\ \bibnamefont {ten Haaf}}, \bibinfo {author} {\bibfnamefont {G.}~\bibnamefont {Wang}}, \bibinfo {author} {\bibfnamefont {N.}~\bibnamefont {van Loo}}, \bibinfo {author} {\bibfnamefont {L.~P.}\ \bibnamefont {Kouwenhoven}},\ and\ \bibinfo {author} {\bibfnamefont {T.}~\bibnamefont {Dvir}},\ }\bibfield  {title} {\bibinfo {title} {Crossed {A}ndreev {R}eflection and {E}lastic {C}otunneling in {T}hree {Q}uantum {D}ots {C}oupled by {S}uperconductors},\ }\href {https://doi.org/10.1103/PhysRevLett.132.056602} {\bibfield  {journal} {\bibinfo  {journal} {Phys. Rev. Lett.}\ }\textbf {\bibinfo {volume} {132}},\ \bibinfo {pages} {056602} (\bibinfo {year}
  {2024}{\natexlab{a}})}\BibitemShut {NoStop}%
\bibitem [{\citenamefont {Zatelli}\ \emph {et~al.}(2024)\citenamefont {Zatelli}, \citenamefont {van Driel}, \citenamefont {Xu}, \citenamefont {Wang}, \citenamefont {Liu}, \citenamefont {Bordin}, \citenamefont {Roovers}, \citenamefont {Mazur}, \citenamefont {van Loo}, \citenamefont {Wolff} \emph {et~al.}}]{zatelli2024robust}%
  \BibitemOpen
  \bibfield  {author} {\bibinfo {author} {\bibfnamefont {F.}~\bibnamefont {Zatelli}}, \bibinfo {author} {\bibfnamefont {D.}~\bibnamefont {van Driel}}, \bibinfo {author} {\bibfnamefont {D.}~\bibnamefont {Xu}}, \bibinfo {author} {\bibfnamefont {G.}~\bibnamefont {Wang}}, \bibinfo {author} {\bibfnamefont {C.-X.}\ \bibnamefont {Liu}}, \bibinfo {author} {\bibfnamefont {A.}~\bibnamefont {Bordin}}, \bibinfo {author} {\bibfnamefont {B.}~\bibnamefont {Roovers}}, \bibinfo {author} {\bibfnamefont {G.~P.}\ \bibnamefont {Mazur}}, \bibinfo {author} {\bibfnamefont {N.}~\bibnamefont {van Loo}}, \bibinfo {author} {\bibfnamefont {J.~C.}\ \bibnamefont {Wolff}}, \emph {et~al.},\ }\bibfield  {title} {\bibinfo {title} {Robust poor man’s {M}ajorana zero modes using {Y}u-{S}hiba-{R}usinov states},\ }\href {https://doi.org/10.1038/s41467-024-52066-2} {\bibfield  {journal} {\bibinfo  {journal} {Nature Communications}\ }\textbf {\bibinfo {volume} {15}},\ \bibinfo {pages} {7933} (\bibinfo {year} {2024})}\BibitemShut {NoStop}%
\bibitem [{\citenamefont {Ten~Haaf}\ \emph {et~al.}(2024)\citenamefont {Ten~Haaf}, \citenamefont {Wang}, \citenamefont {Bozkurt}, \citenamefont {Liu}, \citenamefont {Kulesh}, \citenamefont {Kim}, \citenamefont {Xiao}, \citenamefont {Thomas}, \citenamefont {Manfra}, \citenamefont {Dvir} \emph {et~al.}}]{ten2024two}%
  \BibitemOpen
  \bibfield  {author} {\bibinfo {author} {\bibfnamefont {S.~L.}\ \bibnamefont {Ten~Haaf}}, \bibinfo {author} {\bibfnamefont {Q.}~\bibnamefont {Wang}}, \bibinfo {author} {\bibfnamefont {A.~M.}\ \bibnamefont {Bozkurt}}, \bibinfo {author} {\bibfnamefont {C.-X.}\ \bibnamefont {Liu}}, \bibinfo {author} {\bibfnamefont {I.}~\bibnamefont {Kulesh}}, \bibinfo {author} {\bibfnamefont {P.}~\bibnamefont {Kim}}, \bibinfo {author} {\bibfnamefont {D.}~\bibnamefont {Xiao}}, \bibinfo {author} {\bibfnamefont {C.}~\bibnamefont {Thomas}}, \bibinfo {author} {\bibfnamefont {M.~J.}\ \bibnamefont {Manfra}}, \bibinfo {author} {\bibfnamefont {T.}~\bibnamefont {Dvir}}, \emph {et~al.},\ }\bibfield  {title} {\bibinfo {title} {A two-site {K}itaev chain in a two-dimensional electron gas},\ }\href {https://doi.org/10.1038/s41586-024-07434-9} {\bibfield  {journal} {\bibinfo  {journal} {Nature}\ }\textbf {\bibinfo {volume} {630}},\ \bibinfo {pages} {329} (\bibinfo {year} {2024})}\BibitemShut {NoStop}%
\bibitem [{\citenamefont {Souto}\ \emph {et~al.}(2023)\citenamefont {Souto}, \citenamefont {Tsintzis}, \citenamefont {Leijnse},\ and\ \citenamefont {Danon}}]{Seoane2023}%
  \BibitemOpen
  \bibfield  {author} {\bibinfo {author} {\bibfnamefont {R.~S.}\ \bibnamefont {Souto}}, \bibinfo {author} {\bibfnamefont {A.}~\bibnamefont {Tsintzis}}, \bibinfo {author} {\bibfnamefont {M.}~\bibnamefont {Leijnse}},\ and\ \bibinfo {author} {\bibfnamefont {J.}~\bibnamefont {Danon}},\ }\bibfield  {title} {\bibinfo {title} {Probing {M}ajorana localization in minimal {K}itaev chains through a quantum dot},\ }\href {https://doi.org/10.1103/PhysRevResearch.5.043182} {\bibfield  {journal} {\bibinfo  {journal} {Phys. Rev. Res.}\ }\textbf {\bibinfo {volume} {5}},\ \bibinfo {pages} {043182} (\bibinfo {year} {2023})}\BibitemShut {NoStop}%
\bibitem [{\citenamefont {Alvarado}\ \emph {et~al.}(2024)\citenamefont {Alvarado}, \citenamefont {Yeyati}, \citenamefont {Aguado},\ and\ \citenamefont {Seoane~Souto}}]{Alvarado_PRB2024}%
  \BibitemOpen
  \bibfield  {author} {\bibinfo {author} {\bibfnamefont {M.}~\bibnamefont {Alvarado}}, \bibinfo {author} {\bibfnamefont {A.~L.}\ \bibnamefont {Yeyati}}, \bibinfo {author} {\bibfnamefont {R.}~\bibnamefont {Aguado}},\ and\ \bibinfo {author} {\bibfnamefont {R.}~\bibnamefont {Seoane~Souto}},\ }\bibfield  {title} {\bibinfo {title} {Interplay between {Majorana} and {Shiba} states in a minimal kitaev chain coupled to a superconductor},\ }\href {https://doi.org/10.1103/PhysRevB.110.245144} {\bibfield  {journal} {\bibinfo  {journal} {Phys. Rev. B}\ }\textbf {\bibinfo {volume} {110}},\ \bibinfo {pages} {245144} (\bibinfo {year} {2024})}\BibitemShut {NoStop}%
\bibitem [{\citenamefont {Liu}\ \emph {et~al.}(2023)\citenamefont {Liu}, \citenamefont {Pan}, \citenamefont {Setiawan}, \citenamefont {Wimmer},\ and\ \citenamefont {Sau}}]{liu2023fusion}%
  \BibitemOpen
  \bibfield  {author} {\bibinfo {author} {\bibfnamefont {C.-X.}\ \bibnamefont {Liu}}, \bibinfo {author} {\bibfnamefont {H.}~\bibnamefont {Pan}}, \bibinfo {author} {\bibfnamefont {F.}~\bibnamefont {Setiawan}}, \bibinfo {author} {\bibfnamefont {M.}~\bibnamefont {Wimmer}},\ and\ \bibinfo {author} {\bibfnamefont {J.~D.}\ \bibnamefont {Sau}},\ }\bibfield  {title} {\bibinfo {title} {Fusion protocol for {M}ajorana modes in coupled quantum dots},\ }\href {https://doi.org/10.1103/PhysRevB.108.085437} {\bibfield  {journal} {\bibinfo  {journal} {Phys. Rev. B}\ }\textbf {\bibinfo {volume} {108}},\ \bibinfo {pages} {085437} (\bibinfo {year} {2023})}\BibitemShut {NoStop}%
\bibitem [{\citenamefont {Pandey}\ \emph {et~al.}(2024)\citenamefont {Pandey}, \citenamefont {Okamoto},\ and\ \citenamefont {Dagotto}}]{pandey2024nontrivial}%
  \BibitemOpen
  \bibfield  {author} {\bibinfo {author} {\bibfnamefont {B.}~\bibnamefont {Pandey}}, \bibinfo {author} {\bibfnamefont {S.}~\bibnamefont {Okamoto}},\ and\ \bibinfo {author} {\bibfnamefont {E.}~\bibnamefont {Dagotto}},\ }\bibfield  {title} {\bibinfo {title} {Nontrivial fusion of {M}ajorana zero modes in interacting quantum-dot arrays},\ }\href {https://doi.org/10.1103/PhysRevResearch.6.033314} {\bibfield  {journal} {\bibinfo  {journal} {Phys. Rev. Res.}\ }\textbf {\bibinfo {volume} {6}},\ \bibinfo {pages} {033314} (\bibinfo {year} {2024})}\BibitemShut {NoStop}%
\bibitem [{\citenamefont {Tsintzis}\ \emph {et~al.}(2024)\citenamefont {Tsintzis}, \citenamefont {Souto}, \citenamefont {Flensberg}, \citenamefont {Danon},\ and\ \citenamefont {Leijnse}}]{tsintzis2024majorana}%
  \BibitemOpen
  \bibfield  {author} {\bibinfo {author} {\bibfnamefont {A.}~\bibnamefont {Tsintzis}}, \bibinfo {author} {\bibfnamefont {R.~S.}\ \bibnamefont {Souto}}, \bibinfo {author} {\bibfnamefont {K.}~\bibnamefont {Flensberg}}, \bibinfo {author} {\bibfnamefont {J.}~\bibnamefont {Danon}},\ and\ \bibinfo {author} {\bibfnamefont {M.}~\bibnamefont {Leijnse}},\ }\bibfield  {title} {\bibinfo {title} {Majorana {Q}ubits and {N}on-{A}belian {P}hysics in {Q}uantum {D}ot--{B}ased {M}inimal {K}itaev {C}hains},\ }\href {https://doi.org/10.1103/PRXQuantum.5.010323} {\bibfield  {journal} {\bibinfo  {journal} {PRX Quantum}\ }\textbf {\bibinfo {volume} {5}},\ \bibinfo {pages} {010323} (\bibinfo {year} {2024})}\BibitemShut {NoStop}%
\bibitem [{\citenamefont {Boross}\ and\ \citenamefont {P\'alyi}(2024)}]{boross2024braiding}%
  \BibitemOpen
  \bibfield  {author} {\bibinfo {author} {\bibfnamefont {P.}~\bibnamefont {Boross}}\ and\ \bibinfo {author} {\bibfnamefont {A.}~\bibnamefont {P\'alyi}},\ }\bibfield  {title} {\bibinfo {title} {Braiding-based quantum control of a {M}ajorana qubit built from quantum dots},\ }\href {https://doi.org/10.1103/PhysRevB.109.125410} {\bibfield  {journal} {\bibinfo  {journal} {Phys. Rev. B}\ }\textbf {\bibinfo {volume} {109}},\ \bibinfo {pages} {125410} (\bibinfo {year} {2024})}\BibitemShut {NoStop}%
\bibitem [{\citenamefont {Pino}\ \emph {et~al.}(2024)\citenamefont {Pino}, \citenamefont {Souto},\ and\ \citenamefont {Aguado}}]{pino2024minimal}%
  \BibitemOpen
  \bibfield  {author} {\bibinfo {author} {\bibfnamefont {D.~M.}\ \bibnamefont {Pino}}, \bibinfo {author} {\bibfnamefont {R.~S.}\ \bibnamefont {Souto}},\ and\ \bibinfo {author} {\bibfnamefont {R.}~\bibnamefont {Aguado}},\ }\bibfield  {title} {\bibinfo {title} {Minimal {K}itaev-transmon qubit based on double quantum dots},\ }\href {https://doi.org/10.1103/PhysRevB.109.075101} {\bibfield  {journal} {\bibinfo  {journal} {Phys. Rev. B}\ }\textbf {\bibinfo {volume} {109}},\ \bibinfo {pages} {075101} (\bibinfo {year} {2024})}\BibitemShut {NoStop}%
\bibitem [{\citenamefont {Pan}\ \emph {et~al.}(2025)\citenamefont {Pan}, \citenamefont {Das~Sarma},\ and\ \citenamefont {Liu}}]{pan2024rabi}%
  \BibitemOpen
  \bibfield  {author} {\bibinfo {author} {\bibfnamefont {H.}~\bibnamefont {Pan}}, \bibinfo {author} {\bibfnamefont {S.}~\bibnamefont {Das~Sarma}},\ and\ \bibinfo {author} {\bibfnamefont {C.-X.}\ \bibnamefont {Liu}},\ }\bibfield  {title} {\bibinfo {title} {Rabi and ramsey oscillations of a majorana qubit in a quantum dot-superconductor array},\ }\href {https://doi.org/10.1103/PhysRevB.111.075416} {\bibfield  {journal} {\bibinfo  {journal} {Phys. Rev. B}\ }\textbf {\bibinfo {volume} {111}},\ \bibinfo {pages} {075416} (\bibinfo {year} {2025})}\BibitemShut {NoStop}%
\bibitem [{\citenamefont {Svensson}\ and\ \citenamefont {Leijnse}(2024)}]{svensson2024quantum}%
  \BibitemOpen
  \bibfield  {author} {\bibinfo {author} {\bibfnamefont {V.}~\bibnamefont {Svensson}}\ and\ \bibinfo {author} {\bibfnamefont {M.}~\bibnamefont {Leijnse}},\ }\bibfield  {title} {\bibinfo {title} {Quantum dot based {K}itaev chains: {M}ajorana quality measures and scaling with increasing chain length},\ }\href {https://doi.org/10.1103/PhysRevB.110.155436} {\bibfield  {journal} {\bibinfo  {journal} {Phys. Rev. B}\ }\textbf {\bibinfo {volume} {110}},\ \bibinfo {pages} {155436} (\bibinfo {year} {2024})}\BibitemShut {NoStop}%
\bibitem [{\citenamefont {Luna}\ \emph {et~al.}(2024)\citenamefont {Luna}, \citenamefont {Bozkurt}, \citenamefont {Wimmer},\ and\ \citenamefont {Liu}}]{luna2024flux}%
  \BibitemOpen
  \bibfield  {author} {\bibinfo {author} {\bibfnamefont {J.~D.~T.}\ \bibnamefont {Luna}}, \bibinfo {author} {\bibfnamefont {A.~M.}\ \bibnamefont {Bozkurt}}, \bibinfo {author} {\bibfnamefont {M.}~\bibnamefont {Wimmer}},\ and\ \bibinfo {author} {\bibfnamefont {C.-X.}\ \bibnamefont {Liu}},\ }\bibfield  {title} {\bibinfo {title} {{Flux-tunable Kitaev chain in a quantum dot array}},\ }\href {https://doi.org/10.21468/SciPostPhysCore.7.3.065} {\bibfield  {journal} {\bibinfo  {journal} {SciPost Phys. Core}\ }\textbf {\bibinfo {volume} {7}},\ \bibinfo {pages} {065} (\bibinfo {year} {2024})}\BibitemShut {NoStop}%
\bibitem [{\citenamefont {Escribano}\ \emph {et~al.}(2025)\citenamefont {Escribano}, \citenamefont {Dahl}, \citenamefont {Flensberg},\ and\ \citenamefont {Oreg}}]{escribano2025phasecontrolledminimalkitaevchain}%
  \BibitemOpen
  \bibfield  {author} {\bibinfo {author} {\bibfnamefont {S.~D.}\ \bibnamefont {Escribano}}, \bibinfo {author} {\bibfnamefont {A.~E.}\ \bibnamefont {Dahl}}, \bibinfo {author} {\bibfnamefont {K.}~\bibnamefont {Flensberg}},\ and\ \bibinfo {author} {\bibfnamefont {Y.}~\bibnamefont {Oreg}},\ }\href {https://arxiv.org/abs/2501.14597} {\bibinfo {title} {Phase-controlled minimal kitaev chain in multiterminal josephson junctions}} (\bibinfo {year} {2025}),\ \Eprint {https://arxiv.org/abs/2501.14597} {arXiv:2501.14597 [cond-mat.mes-hall]} \BibitemShut {NoStop}%
\bibitem [{\citenamefont {Miles}\ \emph {et~al.}(2024)\citenamefont {Miles}, \citenamefont {van Driel}, \citenamefont {Wimmer},\ and\ \citenamefont {Liu}}]{Miles2023}%
  \BibitemOpen
  \bibfield  {author} {\bibinfo {author} {\bibfnamefont {S.}~\bibnamefont {Miles}}, \bibinfo {author} {\bibfnamefont {D.}~\bibnamefont {van Driel}}, \bibinfo {author} {\bibfnamefont {M.}~\bibnamefont {Wimmer}},\ and\ \bibinfo {author} {\bibfnamefont {C.-X.}\ \bibnamefont {Liu}},\ }\bibfield  {title} {\bibinfo {title} {Kitaev chain in an alternating quantum dot-{A}ndreev bound state array},\ }\href {https://doi.org/10.1103/PhysRevB.110.024520} {\bibfield  {journal} {\bibinfo  {journal} {Phys. Rev. B}\ }\textbf {\bibinfo {volume} {110}},\ \bibinfo {pages} {024520} (\bibinfo {year} {2024})}\BibitemShut {NoStop}%
\bibitem [{\citenamefont {Souto}\ \emph {et~al.}(2024)\citenamefont {Souto}, \citenamefont {Baran}, \citenamefont {Nitsch}, \citenamefont {Maffi}, \citenamefont {Paaske}, \citenamefont {Leijnse},\ and\ \citenamefont {Burrello}}]{souto2024majorana}%
  \BibitemOpen
  \bibfield  {author} {\bibinfo {author} {\bibfnamefont {R.~S.}\ \bibnamefont {Souto}}, \bibinfo {author} {\bibfnamefont {V.~V.}\ \bibnamefont {Baran}}, \bibinfo {author} {\bibfnamefont {M.}~\bibnamefont {Nitsch}}, \bibinfo {author} {\bibfnamefont {L.}~\bibnamefont {Maffi}}, \bibinfo {author} {\bibfnamefont {J.}~\bibnamefont {Paaske}}, \bibinfo {author} {\bibfnamefont {M.}~\bibnamefont {Leijnse}},\ and\ \bibinfo {author} {\bibfnamefont {M.}~\bibnamefont {Burrello}},\ }\bibfield  {title} {\bibinfo {title} {Majorana modes in quantum dots coupled via a floating superconducting island},\ }\href {https://arxiv.org/abs/2411.07068} {\bibfield  {journal} {\bibinfo  {journal} {arXiv preprint arXiv:2411.07068}\ } (\bibinfo {year} {2024})}\BibitemShut {NoStop}%
\bibitem [{\citenamefont {Seoane~Souto}\ and\ \citenamefont {Aguado}(2024)}]{Souto_arXiv24}%
  \BibitemOpen
  \bibfield  {author} {\bibinfo {author} {\bibfnamefont {R.}~\bibnamefont {Seoane~Souto}}\ and\ \bibinfo {author} {\bibfnamefont {R.}~\bibnamefont {Aguado}},\ }\bibinfo {title} {Subgap {S}tates in {S}emiconductor-{S}uperconductor {D}evices for {Q}uantum {T}echnologies: {A}ndreev {Q}ubits and {M}inimal {M}ajorana {C}hains},\ in\ \href {https://doi.org/10.1007/978-3-031-55657-9_3} {\emph {\bibinfo {booktitle} {New Trends and Platforms for Quantum Technologies}}},\ \bibinfo {editor} {edited by\ \bibinfo {editor} {\bibfnamefont {R.}~\bibnamefont {Aguado}}, \bibinfo {editor} {\bibfnamefont {R.}~\bibnamefont {Citro}}, \bibinfo {editor} {\bibfnamefont {M.}~\bibnamefont {Lewenstein}},\ and\ \bibinfo {editor} {\bibfnamefont {M.}~\bibnamefont {Stern}}}\ (\bibinfo  {publisher} {Springer Nature Switzerland},\ \bibinfo {address} {Cham},\ \bibinfo {year} {2024})\ pp.\ \bibinfo {pages} {133--223}\BibitemShut {NoStop}%
\bibitem [{\citenamefont {Dourado}\ \emph {et~al.}(2025)\citenamefont {Dourado}, \citenamefont {Egues},\ and\ \citenamefont {Penteado}}]{dourado2025twositekitaevsweetspots}%
  \BibitemOpen
  \bibfield  {author} {\bibinfo {author} {\bibfnamefont {R.~A.}\ \bibnamefont {Dourado}}, \bibinfo {author} {\bibfnamefont {J.~C.}\ \bibnamefont {Egues}},\ and\ \bibinfo {author} {\bibfnamefont {P.~H.}\ \bibnamefont {Penteado}},\ }\href {https://arxiv.org/abs/2501.19376} {\bibinfo {title} {Two-site kitaev sweet spots evolving into topological islands}} (\bibinfo {year} {2025}),\ \Eprint {https://arxiv.org/abs/2501.19376} {arXiv:2501.19376 [cond-mat.mes-hall]} \BibitemShut {NoStop}%
\bibitem [{\citenamefont {Bordin}\ \emph {et~al.}(2024{\natexlab{b}})\citenamefont {Bordin}, \citenamefont {Liu}, \citenamefont {Dvir}, \citenamefont {Zatelli}, \citenamefont {ten Haaf}, \citenamefont {van Driel}, \citenamefont {Wang}, \citenamefont {van Loo}, \citenamefont {van Caekenberghe}, \citenamefont {Wolff} \emph {et~al.}}]{bordin2024signatures}%
  \BibitemOpen
  \bibfield  {author} {\bibinfo {author} {\bibfnamefont {A.}~\bibnamefont {Bordin}}, \bibinfo {author} {\bibfnamefont {C.-X.}\ \bibnamefont {Liu}}, \bibinfo {author} {\bibfnamefont {T.}~\bibnamefont {Dvir}}, \bibinfo {author} {\bibfnamefont {F.}~\bibnamefont {Zatelli}}, \bibinfo {author} {\bibfnamefont {S.~L.}\ \bibnamefont {ten Haaf}}, \bibinfo {author} {\bibfnamefont {D.}~\bibnamefont {van Driel}}, \bibinfo {author} {\bibfnamefont {G.}~\bibnamefont {Wang}}, \bibinfo {author} {\bibfnamefont {N.}~\bibnamefont {van Loo}}, \bibinfo {author} {\bibfnamefont {T.}~\bibnamefont {van Caekenberghe}}, \bibinfo {author} {\bibfnamefont {J.~C.}\ \bibnamefont {Wolff}}, \emph {et~al.},\ }\bibfield  {title} {\bibinfo {title} {Signatures of {M}ajorana protection in a three-site {K}itaev chain},\ }\href {https://arxiv.org/abs/2402.19382} {\bibfield  {journal} {\bibinfo  {journal} {arXiv preprint arXiv:2402.19382}\ } (\bibinfo {year} {2024}{\natexlab{b}})}\BibitemShut {NoStop}%
\bibitem [{\citenamefont {ten Haaf}\ \emph {et~al.}(2024)\citenamefont {ten Haaf}, \citenamefont {Zhang}, \citenamefont {Wang}, \citenamefont {Bordin}, \citenamefont {Liu}, \citenamefont {Kulesh}, \citenamefont {Sietses}, \citenamefont {Prosko}, \citenamefont {Xiao}, \citenamefont {Thomas}, \citenamefont {Manfra}, \citenamefont {Wimmer},\ and\ \citenamefont {Goswami}}]{Haaf_arXiv2024}%
  \BibitemOpen
  \bibfield  {author} {\bibinfo {author} {\bibfnamefont {S.~L.~D.}\ \bibnamefont {ten Haaf}}, \bibinfo {author} {\bibfnamefont {Y.}~\bibnamefont {Zhang}}, \bibinfo {author} {\bibfnamefont {Q.}~\bibnamefont {Wang}}, \bibinfo {author} {\bibfnamefont {A.}~\bibnamefont {Bordin}}, \bibinfo {author} {\bibfnamefont {C.-X.}\ \bibnamefont {Liu}}, \bibinfo {author} {\bibfnamefont {I.}~\bibnamefont {Kulesh}}, \bibinfo {author} {\bibfnamefont {V.~P.~M.}\ \bibnamefont {Sietses}}, \bibinfo {author} {\bibfnamefont {C.~G.}\ \bibnamefont {Prosko}}, \bibinfo {author} {\bibfnamefont {D.}~\bibnamefont {Xiao}}, \bibinfo {author} {\bibfnamefont {C.}~\bibnamefont {Thomas}}, \bibinfo {author} {\bibfnamefont {M.~J.}\ \bibnamefont {Manfra}}, \bibinfo {author} {\bibfnamefont {M.}~\bibnamefont {Wimmer}},\ and\ \bibinfo {author} {\bibfnamefont {S.}~\bibnamefont {Goswami}},\ }\href {https://arxiv.org/abs/2410.00658} {\bibinfo {title} {Edge and bulk states in a three-site {K}itaev chain}} (\bibinfo {year} {2024})\BibitemShut {NoStop}%
\bibitem [{\citenamefont {Sticlet}\ \emph {et~al.}(2012)\citenamefont {Sticlet}, \citenamefont {Bena},\ and\ \citenamefont {Simon}}]{sticlet2012spin}%
  \BibitemOpen
  \bibfield  {author} {\bibinfo {author} {\bibfnamefont {D.}~\bibnamefont {Sticlet}}, \bibinfo {author} {\bibfnamefont {C.}~\bibnamefont {Bena}},\ and\ \bibinfo {author} {\bibfnamefont {P.}~\bibnamefont {Simon}},\ }\bibfield  {title} {\bibinfo {title} {Spin and {M}ajorana {P}olarization in {T}opological {S}uperconducting {W}ires},\ }\href {https://doi.org/10.1103/PhysRevLett.108.096802} {\bibfield  {journal} {\bibinfo  {journal} {Phys. Rev. Lett.}\ }\textbf {\bibinfo {volume} {108}},\ \bibinfo {pages} {096802} (\bibinfo {year} {2012})}\BibitemShut {NoStop}%
\bibitem [{\citenamefont {Sedlmayr}\ and\ \citenamefont {Bena}(2015)}]{sedlmayr2015visualizing}%
  \BibitemOpen
  \bibfield  {author} {\bibinfo {author} {\bibfnamefont {N.}~\bibnamefont {Sedlmayr}}\ and\ \bibinfo {author} {\bibfnamefont {C.}~\bibnamefont {Bena}},\ }\bibfield  {title} {\bibinfo {title} {Visualizing {M}ajorana bound states in one and two dimensions using the generalized {M}ajorana polarization},\ }\href {https://doi.org/10.1103/PhysRevB.92.115115} {\bibfield  {journal} {\bibinfo  {journal} {Phys. Rev. B}\ }\textbf {\bibinfo {volume} {92}},\ \bibinfo {pages} {115115} (\bibinfo {year} {2015})}\BibitemShut {NoStop}%
\bibitem [{\citenamefont {Sedlmayr}\ \emph {et~al.}(2016)\citenamefont {Sedlmayr}, \citenamefont {Aguiar-Hualde},\ and\ \citenamefont {Bena}}]{sedlmayr2016majorana}%
  \BibitemOpen
  \bibfield  {author} {\bibinfo {author} {\bibfnamefont {N.}~\bibnamefont {Sedlmayr}}, \bibinfo {author} {\bibfnamefont {J.~M.}\ \bibnamefont {Aguiar-Hualde}},\ and\ \bibinfo {author} {\bibfnamefont {C.}~\bibnamefont {Bena}},\ }\bibfield  {title} {\bibinfo {title} {Majorana bound states in open quasi-one-dimensional and two-dimensional systems with transverse {R}ashba coupling},\ }\href {https://doi.org/10.1103/PhysRevB.93.155425} {\bibfield  {journal} {\bibinfo  {journal} {Phys. Rev. B}\ }\textbf {\bibinfo {volume} {93}},\ \bibinfo {pages} {155425} (\bibinfo {year} {2016})}\BibitemShut {NoStop}%
\bibitem [{\citenamefont {Aksenov}\ \emph {et~al.}(2020)\citenamefont {Aksenov}, \citenamefont {Zlotnikov},\ and\ \citenamefont {Shustin}}]{aksenov2020strong}%
  \BibitemOpen
  \bibfield  {author} {\bibinfo {author} {\bibfnamefont {S.~V.}\ \bibnamefont {Aksenov}}, \bibinfo {author} {\bibfnamefont {A.~O.}\ \bibnamefont {Zlotnikov}},\ and\ \bibinfo {author} {\bibfnamefont {M.~S.}\ \bibnamefont {Shustin}},\ }\bibfield  {title} {\bibinfo {title} {Strong {C}oulomb interactions in the problem of {M}ajorana modes in a wire of the nontrivial topological class {BDI}},\ }\href {https://doi.org/10.1103/PhysRevB.101.125431} {\bibfield  {journal} {\bibinfo  {journal} {Phys. Rev. B}\ }\textbf {\bibinfo {volume} {101}},\ \bibinfo {pages} {125431} (\bibinfo {year} {2020})}\BibitemShut {NoStop}%
\bibitem [{\citenamefont {Wang}\ \emph {et~al.}(2022)\citenamefont {Wang}, \citenamefont {Dvir}, \citenamefont {Mazur}, \citenamefont {Liu}, \citenamefont {van Loo}, \citenamefont {Ten~Haaf}, \citenamefont {Bordin}, \citenamefont {Gazibegovic}, \citenamefont {Badawy}, \citenamefont {Bakkers} \emph {et~al.}}]{wang2022singlet}%
  \BibitemOpen
  \bibfield  {author} {\bibinfo {author} {\bibfnamefont {G.}~\bibnamefont {Wang}}, \bibinfo {author} {\bibfnamefont {T.}~\bibnamefont {Dvir}}, \bibinfo {author} {\bibfnamefont {G.~P.}\ \bibnamefont {Mazur}}, \bibinfo {author} {\bibfnamefont {C.-X.}\ \bibnamefont {Liu}}, \bibinfo {author} {\bibfnamefont {N.}~\bibnamefont {van Loo}}, \bibinfo {author} {\bibfnamefont {S.~L.}\ \bibnamefont {Ten~Haaf}}, \bibinfo {author} {\bibfnamefont {A.}~\bibnamefont {Bordin}}, \bibinfo {author} {\bibfnamefont {S.}~\bibnamefont {Gazibegovic}}, \bibinfo {author} {\bibfnamefont {G.}~\bibnamefont {Badawy}}, \bibinfo {author} {\bibfnamefont {E.~P.}\ \bibnamefont {Bakkers}}, \emph {et~al.},\ }\bibfield  {title} {\bibinfo {title} {Singlet and triplet {C}ooper pair splitting in hybrid superconducting nanowires},\ }\href {https://doi.org/10.1038/s41586-022-05352-2} {\bibfield  {journal} {\bibinfo  {journal} {Nature}\ }\textbf {\bibinfo {volume} {612}},\ \bibinfo {pages} {448} (\bibinfo {year} {2022})}\BibitemShut {NoStop}%
\bibitem [{\citenamefont {Bordoloi}\ \emph {et~al.}(2022)\citenamefont {Bordoloi}, \citenamefont {Zannier}, \citenamefont {Sorba}, \citenamefont {Sch{\"o}nenberger},\ and\ \citenamefont {Baumgartner}}]{Bordoloi_Nature2022}%
  \BibitemOpen
  \bibfield  {author} {\bibinfo {author} {\bibfnamefont {A.}~\bibnamefont {Bordoloi}}, \bibinfo {author} {\bibfnamefont {V.}~\bibnamefont {Zannier}}, \bibinfo {author} {\bibfnamefont {L.}~\bibnamefont {Sorba}}, \bibinfo {author} {\bibfnamefont {C.}~\bibnamefont {Sch{\"o}nenberger}},\ and\ \bibinfo {author} {\bibfnamefont {A.}~\bibnamefont {Baumgartner}},\ }\bibfield  {title} {\bibinfo {title} {Spin cross-correlation experiments in an electron entangler},\ }\href {https://doi.org/10.1038/s41586-022-05436-z} {\bibfield  {journal} {\bibinfo  {journal} {Nature}\ }\textbf {\bibinfo {volume} {612}},\ \bibinfo {pages} {454} (\bibinfo {year} {2022})}\BibitemShut {NoStop}%
\bibitem [{\citenamefont {Awoga}\ and\ \citenamefont {Cayao}(2024)}]{Awoga_PRB24}%
  \BibitemOpen
  \bibfield  {author} {\bibinfo {author} {\bibfnamefont {O.~A.}\ \bibnamefont {Awoga}}\ and\ \bibinfo {author} {\bibfnamefont {J.}~\bibnamefont {Cayao}},\ }\bibfield  {title} {\bibinfo {title} {Identifying trivial and {M}ajorana zero-energy modes using the {M}ajorana polarization},\ }\href {https://doi.org/10.1103/PhysRevB.110.165404} {\bibfield  {journal} {\bibinfo  {journal} {Phys. Rev. B}\ }\textbf {\bibinfo {volume} {110}},\ \bibinfo {pages} {165404} (\bibinfo {year} {2024})}\BibitemShut {NoStop}%
\bibitem [{Note1()}]{Note1}%
  \BibitemOpen
  \bibinfo {note} {The above equation is valid for real Hamiltonians. For $\phi \protect \neq 0$ this is no longer the case, and phase rotations in the $c_{j, \sigma }$ operators are needed to maximize the MP, see Ref.~\cite {svensson2024quantum}}\BibitemShut {NoStop}%
\bibitem [{\citenamefont {Kir{\v{s}}anskas}\ \emph {et~al.}(2017)\citenamefont {Kir{\v{s}}anskas}, \citenamefont {Pedersen}, \citenamefont {Karlstr{\"o}m}, \citenamefont {Leijnse},\ and\ \citenamefont {Wacker}}]{kirvsanskas2017qmeq}%
  \BibitemOpen
  \bibfield  {author} {\bibinfo {author} {\bibfnamefont {G.}~\bibnamefont {Kir{\v{s}}anskas}}, \bibinfo {author} {\bibfnamefont {J.~N.}\ \bibnamefont {Pedersen}}, \bibinfo {author} {\bibfnamefont {O.}~\bibnamefont {Karlstr{\"o}m}}, \bibinfo {author} {\bibfnamefont {M.}~\bibnamefont {Leijnse}},\ and\ \bibinfo {author} {\bibfnamefont {A.}~\bibnamefont {Wacker}},\ }\bibfield  {title} {\bibinfo {title} {Qmeq 1.0: {A}n open-source {P}ython package for calculations of transport through quantum dot devices},\ }\href {https://www.sciencedirect.com/science/article/pii/S0010465517302515} {\bibfield  {journal} {\bibinfo  {journal} {Computer Physics Communications}\ }\textbf {\bibinfo {volume} {221}},\ \bibinfo {pages} {317} (\bibinfo {year} {2017})}\BibitemShut {NoStop}%
\bibitem [{\citenamefont {Zhuo}\ \emph {et~al.}(2025)\citenamefont {Zhuo}, \citenamefont {Yang}, \citenamefont {Huang}, \citenamefont {Lyu}, \citenamefont {Li}, \citenamefont {Li}, \citenamefont {Zhang}, \citenamefont {Wang}, \citenamefont {Wang}, \citenamefont {Shi}, \citenamefont {Wang}, \citenamefont {Bakkers}, \citenamefont {Han}, \citenamefont {Song}, \citenamefont {Li}, \citenamefont {Tong}, \citenamefont {Dou}, \citenamefont {Liu}, \citenamefont {Qu}, \citenamefont {Shen},\ and\ \citenamefont {Lu}}]{zhuo2025readfermionparitypotential}%
  \BibitemOpen
  \bibfield  {author} {\bibinfo {author} {\bibfnamefont {E.}~\bibnamefont {Zhuo}}, \bibinfo {author} {\bibfnamefont {X.}~\bibnamefont {Yang}}, \bibinfo {author} {\bibfnamefont {Y.}~\bibnamefont {Huang}}, \bibinfo {author} {\bibfnamefont {Z.}~\bibnamefont {Lyu}}, \bibinfo {author} {\bibfnamefont {A.}~\bibnamefont {Li}}, \bibinfo {author} {\bibfnamefont {B.}~\bibnamefont {Li}}, \bibinfo {author} {\bibfnamefont {Y.}~\bibnamefont {Zhang}}, \bibinfo {author} {\bibfnamefont {X.}~\bibnamefont {Wang}}, \bibinfo {author} {\bibfnamefont {D.}~\bibnamefont {Wang}}, \bibinfo {author} {\bibfnamefont {Y.}~\bibnamefont {Shi}}, \bibinfo {author} {\bibfnamefont {A.}~\bibnamefont {Wang}}, \bibinfo {author} {\bibfnamefont {E.~P. A.~M.}\ \bibnamefont {Bakkers}}, \bibinfo {author} {\bibfnamefont {X.}~\bibnamefont {Han}}, \bibinfo {author} {\bibfnamefont {X.}~\bibnamefont {Song}}, \bibinfo {author} {\bibfnamefont {P.}~\bibnamefont {Li}}, \bibinfo {author} {\bibfnamefont {B.}~\bibnamefont {Tong}}, \bibinfo {author} {\bibfnamefont
  {Z.}~\bibnamefont {Dou}}, \bibinfo {author} {\bibfnamefont {G.}~\bibnamefont {Liu}}, \bibinfo {author} {\bibfnamefont {F.}~\bibnamefont {Qu}}, \bibinfo {author} {\bibfnamefont {J.}~\bibnamefont {Shen}},\ and\ \bibinfo {author} {\bibfnamefont {L.}~\bibnamefont {Lu}},\ }\href {https://arxiv.org/abs/2501.13367} {\bibinfo {title} {Read out the fermion parity of a potential artificial {K}itaev chain utilizing a transmon qubit}} (\bibinfo {year} {2025}),\ \Eprint {https://arxiv.org/abs/2501.13367} {arXiv:2501.13367 [cond-mat.mes-hall]} \BibitemShut {NoStop}%
\bibitem [{\citenamefont {V\"ayrynen}\ \emph {et~al.}(2015)\citenamefont {V\"ayrynen}, \citenamefont {Rastelli}, \citenamefont {Belzig},\ and\ \citenamefont {Glazman}}]{vayrynen2015microwave}%
  \BibitemOpen
  \bibfield  {author} {\bibinfo {author} {\bibfnamefont {J.~I.}\ \bibnamefont {V\"ayrynen}}, \bibinfo {author} {\bibfnamefont {G.}~\bibnamefont {Rastelli}}, \bibinfo {author} {\bibfnamefont {W.}~\bibnamefont {Belzig}},\ and\ \bibinfo {author} {\bibfnamefont {L.~I.}\ \bibnamefont {Glazman}},\ }\bibfield  {title} {\bibinfo {title} {Microwave signatures of {M}ajorana states in a topological {J}osephson junction},\ }\href {https://doi.org/10.1103/PhysRevB.92.134508} {\bibfield  {journal} {\bibinfo  {journal} {Phys. Rev. B}\ }\textbf {\bibinfo {volume} {92}},\ \bibinfo {pages} {134508} (\bibinfo {year} {2015})}\BibitemShut {NoStop}%
\bibitem [{\citenamefont {Koch}\ \emph {et~al.}(2023)\citenamefont {Koch}, \citenamefont {van Driel}, \citenamefont {Bordin}, \citenamefont {Lado},\ and\ \citenamefont {Greplova}}]{Koch_PRAp2023}%
  \BibitemOpen
  \bibfield  {author} {\bibinfo {author} {\bibfnamefont {R.}~\bibnamefont {Koch}}, \bibinfo {author} {\bibfnamefont {D.}~\bibnamefont {van Driel}}, \bibinfo {author} {\bibfnamefont {A.}~\bibnamefont {Bordin}}, \bibinfo {author} {\bibfnamefont {J.~L.}\ \bibnamefont {Lado}},\ and\ \bibinfo {author} {\bibfnamefont {E.}~\bibnamefont {Greplova}},\ }\bibfield  {title} {\bibinfo {title} {Adversarial {H}amiltonian learning of quantum dots in a minimal {K}itaev chain},\ }\href {https://doi.org/10.1103/PhysRevApplied.20.044081} {\bibfield  {journal} {\bibinfo  {journal} {Phys. Rev. Appl.}\ }\textbf {\bibinfo {volume} {20}},\ \bibinfo {pages} {044081} (\bibinfo {year} {2023})}\BibitemShut {NoStop}%
\bibitem [{\citenamefont {van Driel}\ \emph {et~al.}(2024)\citenamefont {van Driel}, \citenamefont {Koch}, \citenamefont {Sietses}, \citenamefont {ten Haaf}, \citenamefont {Liu}, \citenamefont {Zatelli}, \citenamefont {Roovers}, \citenamefont {Bordin}, \citenamefont {van Loo}, \citenamefont {Wang} \emph {et~al.}}]{van2024cross}%
  \BibitemOpen
  \bibfield  {author} {\bibinfo {author} {\bibfnamefont {D.}~\bibnamefont {van Driel}}, \bibinfo {author} {\bibfnamefont {R.}~\bibnamefont {Koch}}, \bibinfo {author} {\bibfnamefont {V.~P.}\ \bibnamefont {Sietses}}, \bibinfo {author} {\bibfnamefont {S.~L.}\ \bibnamefont {ten Haaf}}, \bibinfo {author} {\bibfnamefont {C.-X.}\ \bibnamefont {Liu}}, \bibinfo {author} {\bibfnamefont {F.}~\bibnamefont {Zatelli}}, \bibinfo {author} {\bibfnamefont {B.}~\bibnamefont {Roovers}}, \bibinfo {author} {\bibfnamefont {A.}~\bibnamefont {Bordin}}, \bibinfo {author} {\bibfnamefont {N.}~\bibnamefont {van Loo}}, \bibinfo {author} {\bibfnamefont {G.}~\bibnamefont {Wang}}, \emph {et~al.},\ }\bibfield  {title} {\bibinfo {title} {Cross-platform {A}utonomous {C}ontrol of {M}inimal {K}itaev {C}hains},\ }\href {https://arxiv.org/abs/2405.04596} {\bibfield  {journal} {\bibinfo  {journal} {arXiv preprint arXiv:2405.04596}\ } (\bibinfo {year} {2024})}\BibitemShut {NoStop}%
\bibitem [{\citenamefont {Benestad}\ \emph {et~al.}(2024)\citenamefont {Benestad}, \citenamefont {Tsintzis}, \citenamefont {Souto}, \citenamefont {Leijnse}, \citenamefont {van Nieuwenburg},\ and\ \citenamefont {Danon}}]{Benestad_PRB2024}%
  \BibitemOpen
  \bibfield  {author} {\bibinfo {author} {\bibfnamefont {J.}~\bibnamefont {Benestad}}, \bibinfo {author} {\bibfnamefont {A.}~\bibnamefont {Tsintzis}}, \bibinfo {author} {\bibfnamefont {R.~S.}\ \bibnamefont {Souto}}, \bibinfo {author} {\bibfnamefont {M.}~\bibnamefont {Leijnse}}, \bibinfo {author} {\bibfnamefont {E.}~\bibnamefont {van Nieuwenburg}},\ and\ \bibinfo {author} {\bibfnamefont {J.}~\bibnamefont {Danon}},\ }\bibfield  {title} {\bibinfo {title} {Machine-learned tuning of artificial {K}itaev chains from tunneling spectroscopy measurements},\ }\href {https://doi.org/10.1103/PhysRevB.110.075402} {\bibfield  {journal} {\bibinfo  {journal} {Phys. Rev. B}\ }\textbf {\bibinfo {volume} {110}},\ \bibinfo {pages} {075402} (\bibinfo {year} {2024})}\BibitemShut {NoStop}%
\bibitem [{Note2()}]{Note2}%
  \BibitemOpen
  \bibinfo {note} {We have verified that the spectral holes vanish as the energy splits or the MP diminishes. However, this is not an exhaustive investigation of non-sweet spots.}\BibitemShut {Stop}%
\bibitem [{\citenamefont {Maiani}\ \emph {et~al.}(2022)\citenamefont {Maiani}, \citenamefont {Geier},\ and\ \citenamefont {Flensberg}}]{maiani2022conductance}%
  \BibitemOpen
  \bibfield  {author} {\bibinfo {author} {\bibfnamefont {A.}~\bibnamefont {Maiani}}, \bibinfo {author} {\bibfnamefont {M.}~\bibnamefont {Geier}},\ and\ \bibinfo {author} {\bibfnamefont {K.}~\bibnamefont {Flensberg}},\ }\bibfield  {title} {\bibinfo {title} {Conductance matrix symmetries of multiterminal semiconductor-superconductor devices},\ }\href {https://doi.org/10.1103/PhysRevB.106.104516} {\bibfield  {journal} {\bibinfo  {journal} {Phys. Rev. B}\ }\textbf {\bibinfo {volume} {106}},\ \bibinfo {pages} {104516} (\bibinfo {year} {2022})}\BibitemShut {NoStop}%
\end{thebibliography}%

\end{document}